\documentclass[oneside,a4paper,11pt,explicit]{book}
\def\ARXIVBUILD{1}
\ifdefined\ARXIVBUILD
\else
  \csname input\endcsname{arxiv_volume_I.tex}
\fi

\usepackage[T1]{fontenc}
\usepackage[utf8]{inputenc}
\usepackage{textcomp}
\usepackage{vol1_kultem}

\usepackage{amsmath,amssymb}
\usepackage{booktabs}
\usepackage{tabularx}
\usepackage{array}
\usepackage{longtable}
\usepackage{lscape}
\usepackage{graphicx}
\usepackage[dvipsnames,svgnames,table]{xcolor}
\usepackage[most]{tcolorbox}
\usepackage{float}
\usepackage{hyperref}

\makeatletter
\renewcommand*\l@section{\@dottedtocline{1}{1.5em}{2.8em}}
\renewcommand*\l@subsection{\@dottedtocline{2}{3.8em}{3.7em}}
\renewcommand*\l@subsubsection{\@dottedtocline{3}{7.0em}{4.6em}}
\makeatother

\usetikzlibrary{arrows.meta,calc,positioning,shapes.geometric,%
  decorations.pathmorphing,backgrounds,fit}
\numberwithin{equation}{chapter}

\definecolor{ink}{HTML}{151B23}
\definecolor{titlebg}{HTML}{100880}
\definecolor{softink}{HTML}{263441}
\definecolor{muted}{HTML}{5C6875}
\definecolor{paper}{HTML}{F6F7F2}
\definecolor{panel}{HTML}{FFFFFF}
\definecolor{line}{HTML}{D8DEE5}
\definecolor{teal}{HTML}{007C77}
\definecolor{blue}{HTML}{245AA6}
\definecolor{gold}{HTML}{B86B00}
\definecolor{rose}{HTML}{A7354D}
\definecolor{green}{HTML}{23724A}
\definecolor{violet}{HTML}{5A4CA0}

\newcolumntype{Y}{>{\raggedright\arraybackslash}X}

\newtcolorbox{leadbox}[2][]{
  enhanced, colback=paper, colframe=#2, boxrule=0.9pt, arc=2mm,
  left=2.2mm, right=2.2mm, top=1.8mm, bottom=1.8mm,
  fonttitle=\sffamily\bfseries, coltitle=white,
  attach boxed title to top left={xshift=2mm,yshift=-2mm},
  boxed title style={colback=#2,arc=1.2mm,boxrule=0pt}, #1 }
\newtcolorbox{metricbox}[1]{
  enhanced, colback=white, colframe=#1, boxrule=0.65pt, arc=1.3mm,
  left=1.8mm, right=1.8mm, top=1.6mm, bottom=1.6mm }

\makeatletter
\renewenvironment{thebibliography}[1]
  {\section*{References}\@mkboth{}{}%
   \list{\@biblabel{\@arabic\c@enumiv}}%
        {\settowidth\labelwidth{\@biblabel{#1}}%
         \leftmargin\labelwidth \advance\leftmargin\labelsep
         \usecounter{enumiv}\let\p@enumiv\@empty
         \renewcommand\theenumiv{\@arabic\c@enumiv}}%
   \small\sloppy\clubpenalty4000\widowpenalty4000\sfcode`\.\@m}
  {\def\@noitemerr{\@latex@warning{Empty `thebibliography' environment}}\endlist}
\makeatother

\hypersetup{colorlinks=true, linkcolor=blue, citecolor=teal, urlcolor=gold,
  pdftitle={3.5-meter Segmented-Mirror Robotic Space Telescope -- Mission White Paper},
  pdfauthor={Sang-Hyun Lee et al.}}

\title{3.5-meter Segmented-Mirror Robotic Space Telescope}
\subtitle{Mission White Paper: I. Overall Architecture and Scientific Mission}
\newcommand{\wpauthors}{Yong-Woo Kang, Sang Hyun Lee, Juhan Kim, Jeong-Yeol Han, Sungwook E. Hong, Bongkon Moon, Donguk Song, Juhyung Kang, Myeong-Gu Park, Sang Chul Kim, Chung-Uk Lee, Sangmo Tony Sohn, Arman Shafieloo, David Parkinson, Hong Soo Park, Dohyeong Kim, Chan Park, Jungjoo Sohn, Young-Beom Jeon, Jong-Hak Woo, Hyung Mok Lee, Hong Bae Ann, Myungkook James Jee, Mansoo Choi, Changbom Park
}
\date{2026}
\newcommand{\wpabstract}{%
We present the preliminary science concept and mission architecture of a 3.5-meter segmented-mirror robotic space telescope currently under study. The telescope is conceived as a versatile space platform capable of supporting several complementary areas of astrophysics, with major programs in wide-field cosmology and galaxy evolution, direct imaging and characterization of nearby planetary systems, time-domain and multi-messenger observations across several programs, compact-object studies, and observations of Solar-System small bodies. These programs place common requirements on angular resolution, photometric stability, rapid target acquisition, spectroscopic capability, and long-term observing efficiency.
The telescope employs a segmented 3.5-meter primary mirror and is designed for high-angular-resolution imaging from the near-ultraviolet through the optical and near-infrared. The current baseline covers 0.2--1.5~$\mu$m, with the wavelength at which diffraction-limited performance is specified to be determined from the final wavefront-error budget.
A wide-field imaging system is intended to support deep surveys, precision photometry, and repeated monitoring over a field of view of approximately $10' \times 10'$ to $30' \times 30'$. Spectroscopic modes with resolving powers around $R \sim 1000$, together with higher-resolution options approaching $R \sim 5000$, are being considered for galaxy surveys, transient classification, compact-object spectroscopy, and other targeted investigations.
A dedicated coronagraphic capability is also being studied for direct observations of nearby exoplanetary systems. The current performance goal is a raw contrast of order $10^{-8}$, with further improvement expected through calibration and post-processing, while maintaining the angular resolution required to probe planetary systems around the nearest stars. Candidate mission configurations include operations near the Sun--Earth L2 region as well as alternative Earth-orbit architectures, with the final choice to be determined by science performance, thermal stability, communications, operations, and mission cost.
The telescope is intended to combine survey observations, long-duration monitoring, and time-sensitive observations. The telescope integrates wide-field imaging, spectroscopy, and 
coronagraphic capability around a common 18-segment optical system.
This paper defines the current science requirements, baseline technical
configuration, and engineering trade space for subsequent development
of the 3.5mST concept.
}

\newcommand{\wpkeywords}{\textbf{Keywords:} segmented mirror, robotic space telescope, cosmology, galaxy evolution, exoplanet, time-domain astronomy, multi-messenger astronomy}

\newcommand{\MakeFrontCover}{%
  \begin{titlepage}
  \thispagestyle{empty}\sffamily\centering
  \noindent\colorbox{Blue3}{\parbox[t]{\dimexpr\textwidth-2\fboxsep\relax}{%
    \vspace{4mm}\centering
    {\color{white}\bfseries\fontsize{24}{29}\selectfont 3.5-meter Segmented-Mirror\\[1.5mm]
      Robotic Space Telescope}\\[3.5mm]
    {\color{white}\Large Mission White Paper}\\[1.5mm]
    {\color{white!88}\large I. Overall Architecture and Scientific Mission}%
    \vspace{4mm}}}
  \vfill
  \includegraphics[width=\textwidth]{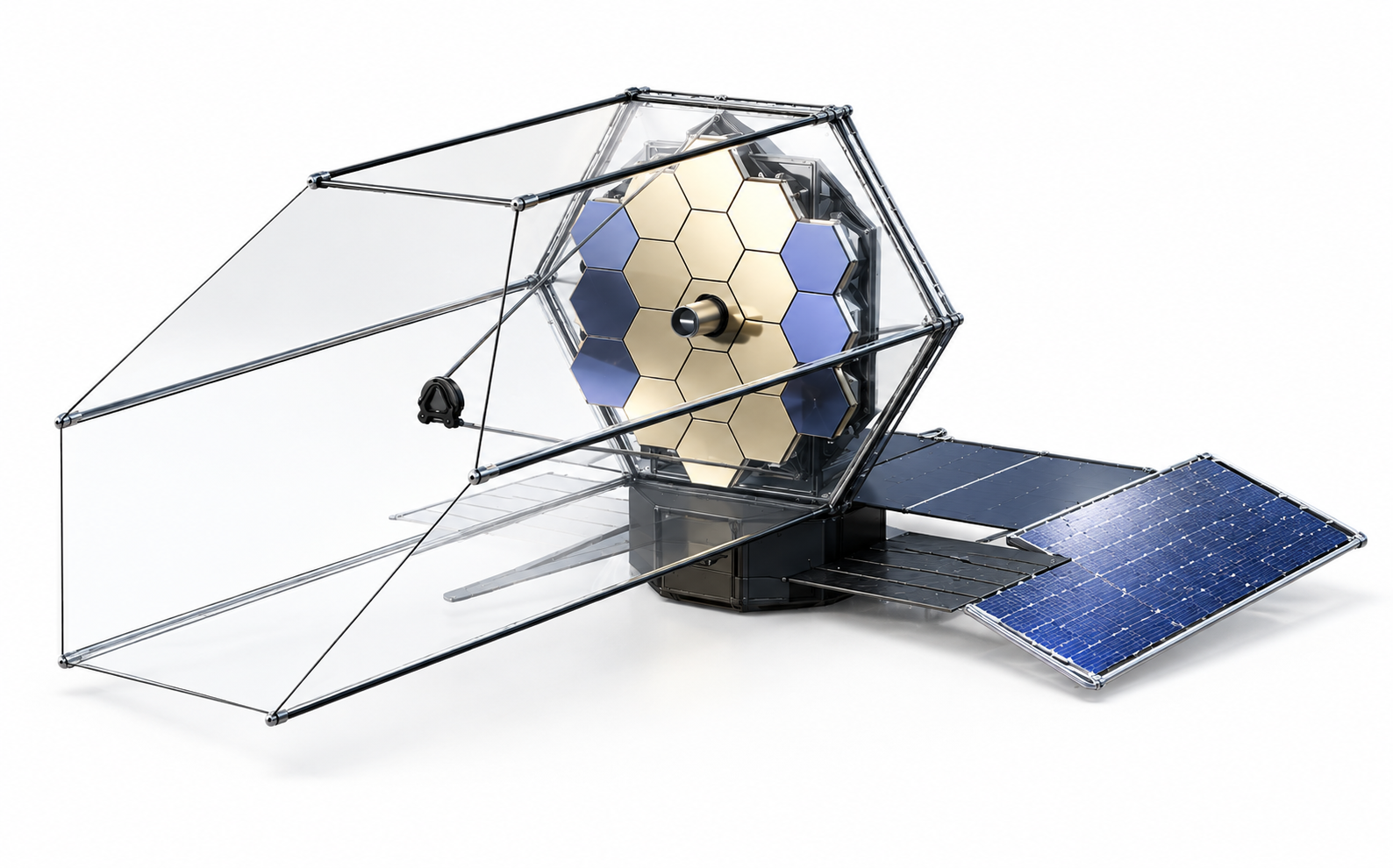}
  \vfill
  {\color{ink}\bfseries\normalsize \wpauthors\par}
  \vspace{3mm}
  {\color{muted}\small 
    Korea Astronomy and Space Science Institute \textperiodcentered\   
    University of Ulsan \\ 
    University of Science and Technology \textperiodcentered\
    Kyungpook National University\\
    Space Telescope Science Institute \textperiodcentered\ 
    Korea Institute for Advanced Study \\ 
    Pusan National University \textperiodcentered\ 
    Korea National University of Education\\
    Seoul National University \textperiodcentered\
    Yonsei University
    \par}
    
  \vspace{5mm}
  {\color{Blue3}\bfseries\Large 2026}
  \vspace{4mm}
  \end{titlepage}}

\newcommand{\MakeColophon}{%
  \clearpage
  \thispagestyle{empty}\sffamily
  \null\vfill
  \noindent{\color{muted}\footnotesize  
    {\color{ink}\bfseries 3.5-meter Segmented-Mirror Robotic Space Telescope}\\
    Mission White Paper: I. Overall Architecture and Scientific Mission\\[5mm]
    {\color{rose}\bfseries Release date September 17, 2026 \\ \\}
    \textcopyright\ 2026 The 3.5mST Mission Study Team and the authors. All rights reserved.
    \\[2mm]
    Prepared by a community-based mission study team consisting of astronomers and researchers from multiple institutions in Korea and abroad.
    \\[5mm]    
    {\color{ink}\bfseries Concept Status:}\ The 3.5mST is currently at an early concept-study stage. Mission funding has not yet been secured, and formal mission development has not begun. This white paper summarizes initial scientific and technical discussions and is intended to introduce the concept to a broader community, invite further input, and provide a basis for subsequent discussion and development.\\[2mm]    
    {\color{ink}\bfseries Suggested citation:}\ Kang, Y.-W., Lee, S.-H., et
    al.\ (2026), \textit{3.5-meter Segmented-Mirror Robotic Space Telescope:
    Mission White Paper}.\\[2mm]
    {\color{ink}\bfseries Corresponding author:}\ Sang Hyun Lee
    \textperiodcentered\ \texttt{shlee@kasi.re.kr}.\\[5mm]
    {\color{ink}\bfseries AI-Assisted Illustrations:}\ Generative-AI tools were used solely for illustrative purposes in selected images in this white paper. AI-generated illustrations include the small illustrative images on pp. ii--iv and Figures 4.1, 4.3 (right panel), 4.4, and 4.5. The cover illustration was originally created by a human designer and subsequently retouched using AI-assisted image tools. Generative AI was not used to produce scientific data, observational results, quantitative simulations, or scientific measurements presented in this white paper. All externally sourced scientific images and data are credited to their original sources where applicable.
    \\[5mm]
    
    \par}
  \vspace{8mm}
  \clearpage}

\newcommand{\MakeBackCover}{%
  \thispagestyle{empty}\sffamily\centering
  
  \null\vspace*{\stretch{1}}
  \includegraphics[width=\textwidth]{vol1_3.5ST_GPT.png}
  \vspace*{\stretch{2.2}}
  \noindent\colorbox{Blue3}{\parbox[t]{\dimexpr\textwidth-2\fboxsep\relax}{%
    \vspace{3mm}\centering
    {\color{white}\bfseries\Large 3.5-meter Segmented-Mirror Robotic Space Telescope}\\[1.5mm]
    {\color{white!88}\normalsize Mission White Paper: I. Overall Architecture and Scientific Mission}%
    \vspace{3mm}}}
  \vspace{3mm}
  \clearpage}

\begin{document}
\frontmatter
\MakeFrontCover
%\MakeAuthorPage
\MakeColophon        % inside front cover: copyright / license / citation

% ===== Mission Architecture infographic =====
\clearpage
%=========================================================
% Executive Summary Figures
%=========================================================

%-------------------------------
% Figure 1
%-------------------------------
\clearpage
\thispagestyle{fancy}

\begin{center}
\vspace*{0.0cm}
\includegraphics[
    width=1.0\textwidth,
    height=\textheight,
    keepaspectratio
]{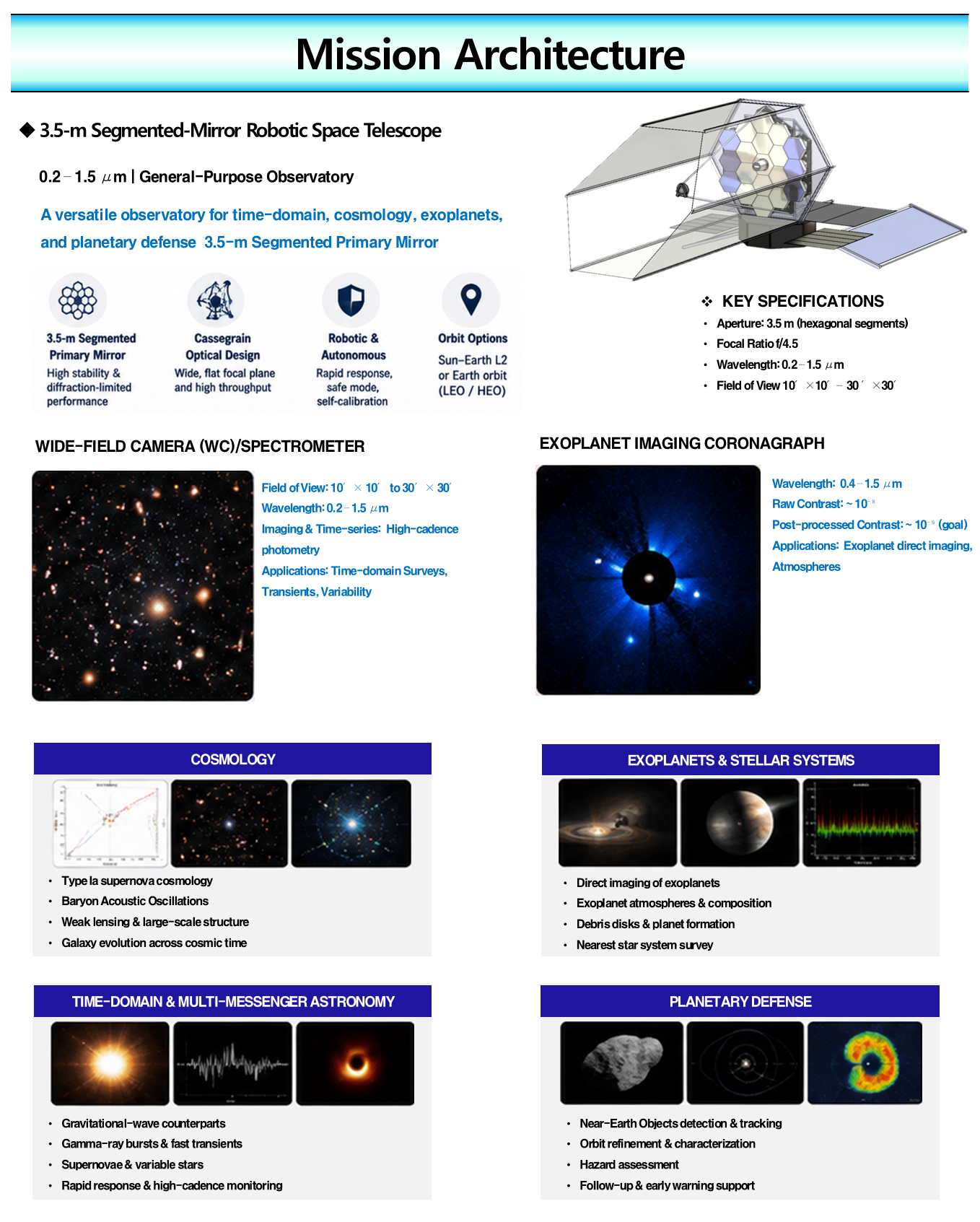}
\end{center}

%-------------------------------
% Figure 2
%-------------------------------
\clearpage
\thispagestyle{fancy}

\begin{center}
\vspace*{0.0cm}
\includegraphics[
    width=1.0\textwidth,
    height=\textheight,
    keepaspectratio
]{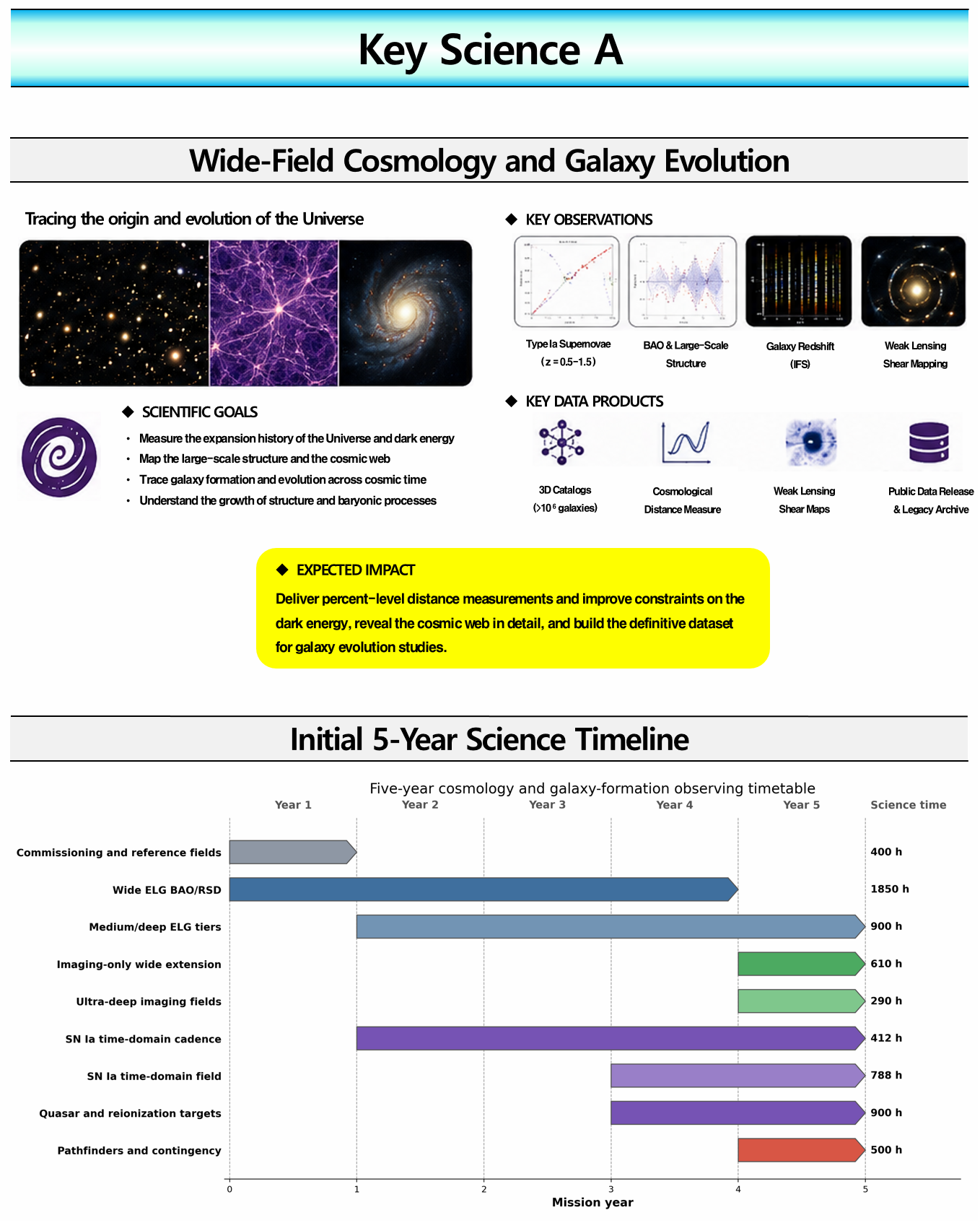}
\end{center}

%-------------------------------
% Figure 3
%-------------------------------
\clearpage
\thispagestyle{fancy}

\begin{center}
\vspace*{0.0cm}
\includegraphics[
    width=1.0\textwidth,
    height=\textheight,
    keepaspectratio
]{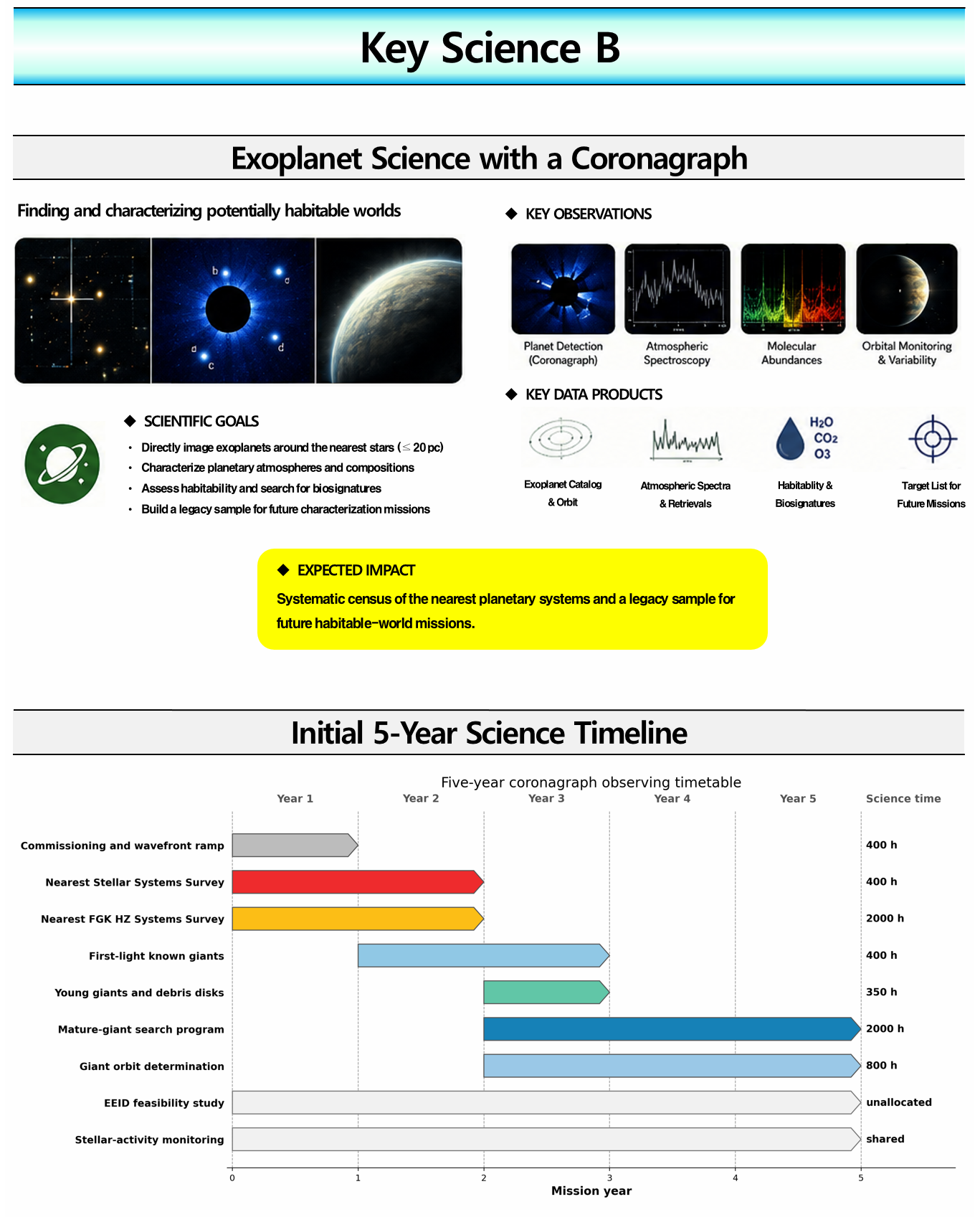}
\end{center}

\clearpage
\clearpage

\thispagestyle{fancy}
\vspace*{0.5em}
\noindent{\Large\bfseries Abstract}\par\smallskip
\noindent\wpabstract\par\medskip
\noindent\wpkeywords
\clearpage

\tableofcontents

\mainmatter
\chapter{Scientific Context and Mission Rationale}
\label{chap:introduction}

Astronomy in the 2030s will require space observatories capable of addressing scientific problems that span very different spatial and temporal scales. Precision cosmology and galaxy studies require stable imaging and spectroscopy over repeated observations, compact-object and transient science benefits from flexible time-domain measurements, and the characterization of nearby planetary systems demands high angular resolution and precise control of stellar light. The proposed 3.5-meter Segmented-Mirror Robotic Space Telescope (3.5mST) is being investigated as a community-driven concept that brings these complementary requirements together within a common optical and near-infrared observing platform.

A defining feature of the 3.5mST concept is the use of a segmented primary mirror to obtain the
collecting area and angular resolution of a 3.5-meter-class telescope within realistic launch constraints. This architecture makes segment phasing, wavefront sensing and control, and long-term optical stability central elements of the mission from the beginning of the concept study. At the same time, the telescope is intended to support wide-field imaging, spectroscopy, repeated photometric monitoring, and coronagraphic observations without being optimized exclusively for any single science program. The mission study therefore begins with the observational requirements shared across its principal science areas and examines how they can be accommodated within a technically coherent segmented-mirror system.

The broader scientific context reinforces the value of such a multi-purpose space observatory.
Gravitational-wave events such as GW170817 have demonstrated the importance of coordinated
electromagnetic observations of compact-object mergers \cite{abbott2017}, while rapidly evolving transients continue to motivate repeated observations over short timescales  \cite{margutti2019}. Wide-field facilities such as the Vera C. Rubin Observatory are expanding the discovery space for galaxies, variable sources, and transient phenomena \cite{ivezic2019}, and the Nancy Grace Roman Space Telescope illustrates the scientific reach of stable wide-field observations from space \cite{roman2026}. 

In developing the present white paper, we also consulted the Lazuli Space Observatory studies as contemporary references for optical and near-infrared space-observatory concepts \cite{roy2026,wevers2026}. These studies provide useful comparison points in areas such as time-domain observations, spectroscopy, and high-contrast imaging, while the 3.5mST is developed around its own science priorities, segmented-mirror architecture, and associated phasing and wavefront-control requirements.

An important component of the 3.5mST science case is the direct investigation of planetary systems around nearby stars. Direct imaging provides access to planetary light that is spatially separated from the host star and can therefore constrain the architectures and physical properties of planetary systems in ways complementary to transit and radial-velocity measurements \cite{zurlo2024}. Achieving this capability from space requires not only sufficient aperture, but also precise wavefront control, stable pointing, and dedicated suppression of stellar light. For this reason, coronagraphic observations are considered
together with the telescope optical architecture from the present concept stage.

The 3.5mST concept is built around an 18-segment primary mirror with an
effective aperture of approximately 3.5~m. The segmented architecture is
intended to provide the collecting area and angular resolution required
for the science program while remaining compatible with launch constraints.
Active segment phasing and wavefront control are therefore fundamental
technical elements of the telescope concept. 
Existing Korean experience in segmented-mirror phasing and integration,
including technology development carried out at KASI, provides a useful
technical basis for evaluating the feasibility of the proposed space-based
segmented-mirror system.

The current science concept combines three principal observational
capabilities. Wide-field imaging is intended to support deep surveys,
precision photometry, and repeated monitoring over a field of view of
approximately $10' \times 10'$ to $30' \times 30'$. Spectroscopy is being
studied with a planning baseline of approximately $R \sim 1000$, together
with higher-resolution options approaching $R \sim 5000$ for science cases
requiring resolved spectral structure. A coronagraphic mode is included
for investigations of nearby planetary systems, with a current raw-contrast
goal of order $10^{-8}$. These capabilities are supported by an optical
and near-infrared baseline wavelength range of 0.2--1.5~$\mu$m, while the
scientific and technical value of extending calibrated performance toward
longer near-infrared wavelengths remains under study.

These observing capabilities are intended to serve several scientific
programs rather than a single experiment. The principal areas considered
in this white paper include cosmology and galaxy evolution, compact-object
astrophysics, time-domain and multi-messenger phenomena, direct imaging
and characterization of exoplanetary systems, and observations of
Solar-System small bodies. Their requirements differ substantially:
cosmological surveys favor wide field and well-controlled photometry and
spectroscopy; compact-object and transient studies benefit from repeated
imaging and spectroscopy; and direct exoplanet observations impose stringent
requirements on wavefront stability and stellar-light suppression.
Consequently, the present mission study focuses on identifying a telescope
configuration that can satisfy these requirements within a technically
realistic common architecture.

The 3.5mST concept is currently at a very early stage of development. No mission funding has yet been secured, and formal mission development has not begun. The work to date has primarily consisted of scientific and technical brainstorming among researchers to identify possible science goals, telescope capabilities, and mission concepts. This white paper represents an effort to organize those discussions into an initial science and architecture concept, to invite broader input from the Korean astronomical community, and to introduce the concept to a wider audience as a basis for further discussion.

Accordingly, the present white paper should not be regarded as a finalized mission design or an approved mission plan. Several important parameters, including the detailed instrument configuration, near-infrared wavelength extension, final orbit, detector architecture, and achievable coronagraph performance, remain subjects of continuing trade studies. The purpose of this study is therefore to define the scientific requirements, establish an initial technically coherent baseline, and identify the areas that require further quantitative analysis and technology development.

This paper is organized as follows. Chapter~2 summarizes the scientific
motivation in the principal astrophysical areas considered for the mission.
Chapter~3 presents the baseline telescope configuration and the technical
requirements associated with the segmented primary mirror, wavefront
control, imaging, spectroscopy, coronagraphy, spacecraft, and mission
environment. Chapter~4 develops the principal scientific objectives in
greater detail. Chapter~5 summarizes the resulting science and technology
concept and identifies the next steps required to refine the 3.5mST
mission.

%%%%%%%%%%%%%%%%%%
\chapter{Science Motivation}

The science case of the 3.5-meter Segmented-Mirror Robotic Space Telescope
is organized around the physical quantities that the telescope is expected
to measure. These include photometric variability, spectral features and
redshifts, angular and spatial structure, astrometric motion, and faint
signals in the vicinity of bright sources. The combination of these
measurements defines the scientific role of the telescope more directly
than any single observing mode or target class.

In this white paper, several research areas are considered. Cosmology uses
photometric and spectroscopic measurements to study cosmic expansion,
large-scale structure, and the distribution of matter. Galaxy studies
require spatially resolved imaging and spectroscopy of stellar populations,
gas, and nuclear activity. Compact-object science makes use of repeated
photometry and spectroscopy to investigate stellar remnants, accretion,
dense matter, and variable phenomena. Exoplanet science combines precision
photometry, spectroscopy, and high-contrast imaging to study planetary
systems and their environments. Near-Earth Object research relies primarily
on accurate astrometry, time-resolved photometry, and visible/near-infrared
spectrophotometry to determine orbital and physical properties.

These science cases provide the basis for defining the observational
requirements of 3.5mST. Rather than treating the telescope capabilities as
independent design goals, the following sections identify how each
measurement contributes to a specific astrophysical question. The detailed
science programs and their quantitative objectives are developed further in
Chapter~4.

\section{Cosmology}

The cosmology program of the 3.5-meter Segmented-Mirror Robotic Space
Telescope is motivated by the need to measure the expansion history of the
Universe and to test the consistency of cosmological parameters using
multiple observational probes. The discovery of cosmic acceleration through
Type Ia supernovae established dark energy as a central component of the
standard cosmological model, while its physical origin remains unresolved
\cite{riess98,perlmutter99,planck2020}.

For 3.5mST, the relevant cosmological measurements are those that benefit
from stable and repeatable optical and near-infrared observations. These
include multi-epoch photometry and spectroscopy of Type Ia supernovae,
redshift measurements of galaxies, and imaging data that can support
gravitational-lensing studies. The combination of these measurements enables
independent estimates of cosmic distances, expansion history, and the growth
and distribution of matter.

A particular emphasis is placed on controlling observational systematics
through homogeneous measurements obtained with the same space-based
instrumentation over repeated epochs. For Type Ia supernovae, this approach
can improve the characterization of color evolution, spectral properties,
and population-dependent effects that enter cosmological distance
measurements. Galaxy spectroscopy and imaging provide complementary
information on large-scale structure and matter distribution.

Electromagnetic observations of gravitational-wave sources can provide an
additional cosmological constraint when suitable counterparts are identified.
In the 3.5mST science concept, such observations are treated as one component
of a broader cosmology program rather than as the defining motivation of the
mission. Together, supernova, galaxy, lensing, and multi-messenger
measurements provide a set of complementary tests of cosmic expansion and
the underlying cosmological model.

\section{Galaxies}

The formation and evolution of galaxies result from the complex interplay of star formation, gas accretion and circulation, and the growth of supermassive black holes. These processes operate over a wide range of timescales throughout cosmic history and leave observable signatures in galaxy morphology, color, chemical composition, and nuclear activity. A central goal of galaxy evolution studies is to establish the physical connections between these observable properties and the underlying astrophysical processes that drive them.

The optical and near-infrared wavelength ranges provide the most direct means of tracing stellar populations, interstellar gas, and galactic structures. However, ground-based observations are limited by atmospheric emission, absorption, and seeing effects, which hinder precise measurements of fine structural features and faint low-surface-brightness components. In particular, systematic investigations of galaxy outskirts, diffuse low-surface-brightness structures, and the complex environments surrounding galactic nuclei are difficult without space-based observations.

With its near-diffraction-limited spatial resolution and stable spectroscopic performance, the 3.5-meter Segmented-Mirror Robotic Space Telescope would enable detailed studies of the structure and physical properties of nearby and intermediate-redshift galaxies \cite{springel2005}. By simultaneously tracing star-forming regions, gas distributions, and the variability characteristics of active galactic nuclei (AGNs), the telescope would provide direct observational constraints on the energy feedback mechanisms that regulate galaxy evolution. In particular, time-domain observations of AGN variability would offer valuable insights into the connection between black hole growth and the evolutionary pathways of galaxies.

Furthermore, the telescope would play a key role in international collaborative research by providing deeper and more precise follow-up observations of targets selected from large-scale ground-based survey programs. Such capabilities would enhance both the visibility and scientific leadership of Korean researchers in the field of galaxy evolution studies.

\section{Compact Objects}

Compact objects provide access to physical regimes that cannot be reproduced under ordinary astrophysical conditions. White dwarfs, neutron stars, and black holes probe matter at extreme density, strong gravitational fields, intense magnetic environments, and highly energetic accretion processes. Observations of these systems therefore offer complementary tests of stellar evolution, dense-matter physics, accretion and outflow mechanisms, and the production and transport of energy in relativistic environments.

The compact-object program of the 3.5-meter Segmented-Mirror Robotic Space Telescope is built around both persistent and variable sources. High-precision time-series photometry of white dwarfs can be used to study stellar pulsations and internal structure, while repeated spectroscopic observations of cataclysmic variables and compact binaries can trace changes in accretion state, orbital phase, and circumstellar material. Spectroscopic monitoring of neutron-star and black-hole systems would provide additional diagnostics of line profiles, velocity structure, and the physical conditions of matter interacting with compact objects.

Multi-messenger sources constitute another important component of this program.
The detection of GW170817 demonstrated the scientific importance of compact-object
mergers as multi-messenger sources \cite{abbott2017}. For future gravitational-wave
events and other externally triggered phenomena, the telescope can carry out imaging
and spectroscopic follow-up to characterize their electromagnetic counterparts and
subsequent evolution. Such observations can connect the properties of compact-object
systems inferred from gravitational-wave measurements with independently measured
electromagnetic emission, contributing to studies of merger physics, heavy-element
production, and the evolution of merger remnants.

These investigations place complementary demands on the telescope: stable long-duration photometry, repeatable spectroscopy over multiple epochs, flexible scheduling, and the capability to accommodate time-sensitive observations when scientifically warranted. By combining these observing modes within a single space-based facility, the 3.5-meter telescope would support compact-object studies across timescales ranging from rapid variability to long-term evolutionary changes and would provide a common observational platform for stellar-remnant physics, accretion studies, and multi-messenger astrophysics.

\section{Exoplanets}

The exoplanet program of the 3.5-meter Segmented-Mirror Robotic Space
Telescope is centered on the physical characterization of planetary systems
through photometric, spectroscopic, and high-contrast observations. Current
exoplanet research increasingly focuses not only on the detection of planets,
but also on understanding their atmospheric properties, system architectures,
and interactions with their host stars \cite{seager2013}.

Precision photometry and spectroscopy can be used to investigate planetary
atmospheres and to measure changes associated with transits, orbital phase,
and stellar variability. Repeated observations are particularly important
for separating planetary signals from time-dependent stellar activity and
for improving the consistency of atmospheric measurements across multiple
epochs.

A complementary component of the program is the direct observation of nearby
planetary systems with a stellar coronagraph. High-contrast imaging can
provide spatially resolved measurements of planets and circumstellar
environments that are inaccessible through transit observations alone.
For favorable nearby systems, repeated coronagraphic observations can be
used to constrain projected orbital motion, relative brightness, and the
wavelength dependence of planetary reflected light. Spectroscopic follow-up
of sufficiently bright directly imaged targets can further provide
information on atmospheric composition and cloud properties.

These measurements address several related scientific questions: how common
different planetary-system architectures are, how planetary properties vary
with host-star characteristics, how stellar activity influences observable
planetary environments, and which nearby systems provide the most favorable
conditions for detailed characterization. The detailed observing strategies,
performance limits, and target classes considered for these investigations
are presented in Chapter~4.

\section{Near-Earth Objects}

Near-Earth Objects (NEOs) are remnants of the formation and evolution of the Solar System and, at the same time, represent potential impact hazards to Earth \cite{mainzer2011}. Accurate determination of their orbital parameters and physical properties is therefore important not only for scientific investigations but also for planetary defense and public safety.

Ground-based survey programs have successfully discovered a large number of NEOs; however, refining their orbital elements and characterizing their physical properties often require additional space-based observations. Repeated astrometric measurements with the 3.5-meter Segmented-Mirror Robotic Space Telescope would improve orbit determination and long-term trajectory predictions, while spectroscopic observations would constrain the composition, taxonomy, and surface properties of these objects \cite{nasa2023}.

The mission concept envisages close coordination with international planetary-defense networks, playing a critical role throughout the entire process of discovery, tracking, and physical characterization of potentially hazardous objects. As such, the telescope would represent a compelling example of how a Korean space telescope can contribute not only to fundamental scientific research but also directly to the safety and well-being of human society.

\chapter{Baseline Telescope Configuration and Technical Requirements}

\section{Baseline Mission Configuration}

The 3.5-meter Segmented-Mirror Robotic Space Telescope is being studied as
a large-aperture optical and near-infrared telescope that combines survey
capability, precision time-series observations, spectroscopy, and
high-contrast imaging within a common telescope platform. The telescope
concept is defined by the scientific requirements described in the preceding
chapter rather than by a single observing mode.

The current baseline adopts a 3.5-meter segmented primary mirror consisting
of 18 hexagonal mirror segments. The telescope is configured to provide
high angular resolution and stable imaging performance across the optical
and near-infrared wavelength range, while supporting several scientific
instruments with different requirements for field of view, spectral
resolution, cadence, and wavefront stability.

The baseline wavelength range is 0.2--1.5~$\mu$m. Extension of calibrated
throughput toward longer near-infrared wavelengths remains under
investigation and would be evaluated through subsequent optical, detector,
thermal, and sensitivity studies. The wide-field imaging requirement spans
approximately $10' \times 10'$ to $30' \times 30'$, depending on the final
optical and detector configuration.

The principal instrument capabilities currently considered for the
telescope are wide-field imaging, spectroscopy, and coronagraphic imaging.
Time-series photometry and high-cadence monitoring are supported primarily
through the wide-field imaging system.

\begin{table}[ht]
\centering
\caption{Current baseline parameters of the 3.5-meter Segmented-Mirror
Robotic Space Telescope. Values remain subject to refinement during
subsequent design studies.}
\begin{tabular}{ll}
\hline
Parameter & Current Baseline \\
\hline
Primary mirror diameter & 3.5~m \\
Primary mirror architecture & 18 hexagonal mirror segments \\
Optical configuration & On-axis segmented-mirror telescope \\
Focal ratio & $f/4.5$ \\
Baseline wavelength range & 0.2--1.5~$\mu$m \\
Field of view & $10' \times 10'$ to $30' \times 30'$ \\
Imaging & Wide-field near-UV/optical/near-infrared imaging \\
Spectroscopy & $R \sim 1000$ baseline; $R \sim 5000$ option \\
High-contrast capability & Stellar coronagraph \\
Mission orbit & Under study: Sun--Earth L2 or Earth orbit \\
Baseline mission lifetime & 10 years \\
Launch constraint & Approximately 3~m fairing diameter \\
\hline
\end{tabular}
\end{table}

%%%%%%%%%%%%%%%%%%%%%%%%%%%%
\section{Segmented Primary Mirror and Phasing Architecture}
\begin{figure}[htbp]
  \centering
  \includegraphics[width=1.5\linewidth,height=0.75\textheight,keepaspectratio]{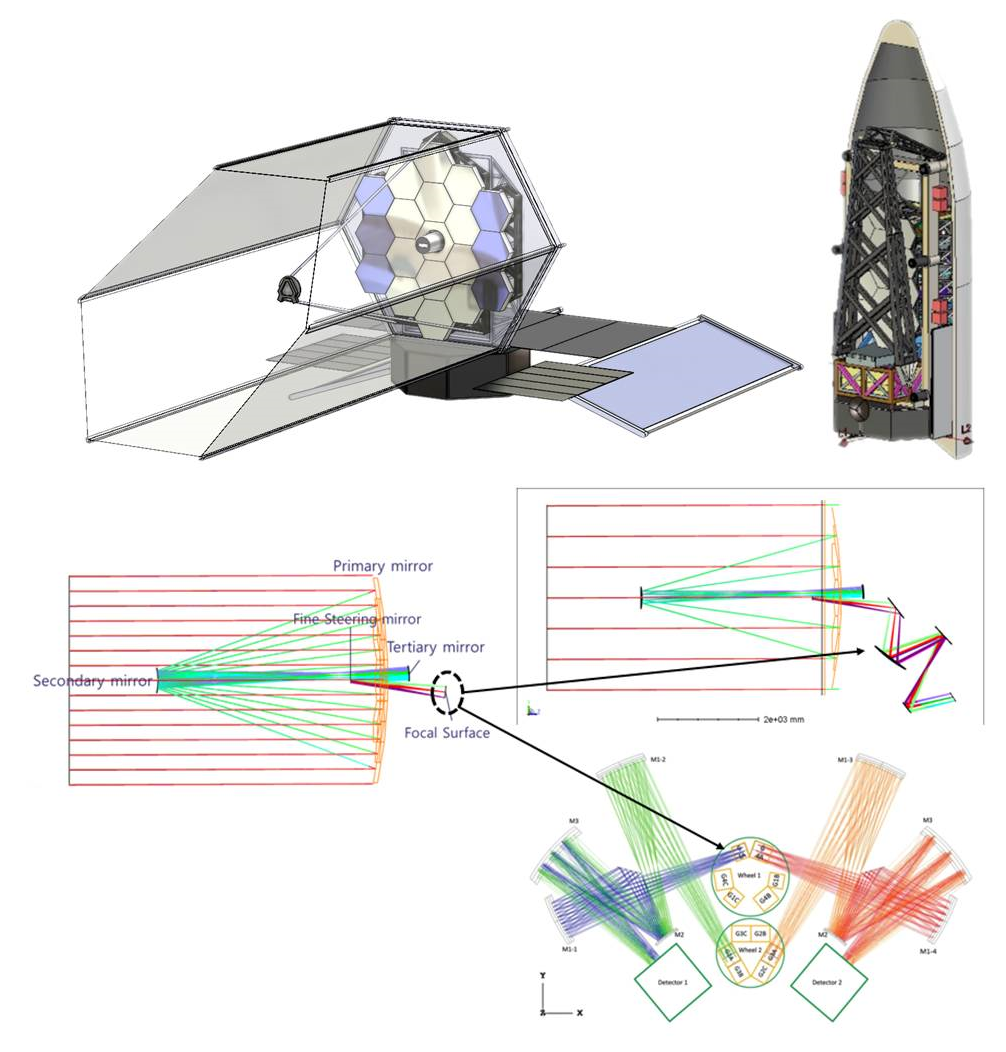}
  \caption{Schematic diagram of the 3.5-meter segmented-mirror robotic space
  telescope and its ultraviolet, visible, and near-infrared instrument concepts}
  \label{fig:telescope_overview}
\end{figure}
%%%%%%%%%%%%%%%%%%%%%%%%%%%%%%%

\vspace{8mm}
\begin{figure}[htbp]
  \centering
  \includegraphics[width=0.91\linewidth,height=0.46\textheight,keepaspectratio]{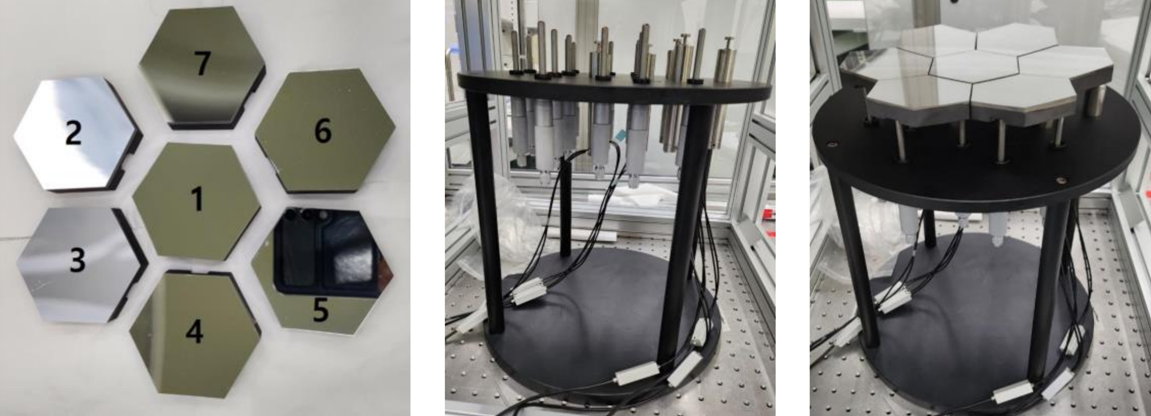}
  \caption{Demonstration of segmented-mirror phasing technology developed
  by KASI.}
  \label{fig:kasi-segment-phasing}
\end{figure}
\vspace{8mm}

\begin{figure}[htbp]
  \centering
  \includegraphics[width=0.91\linewidth,height=0.46\textheight,keepaspectratio]{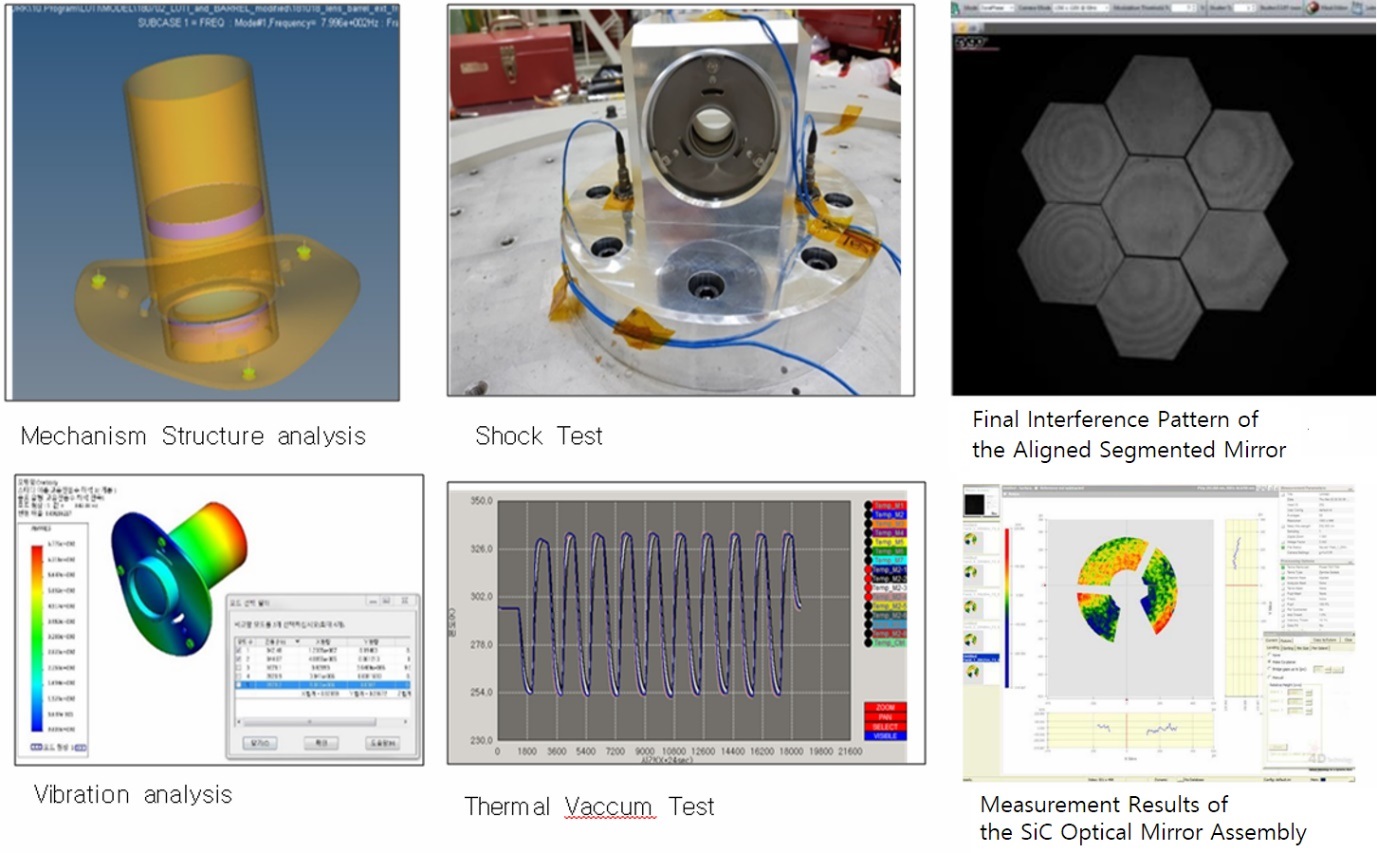}
  \caption{Attachment of fabricated mirror segments to the support
  structure, installation of actuators and micrometers, and alignment and
  control using an interferometer.}
  \label{fig:kasi-segment-integration}
\end{figure}
\vspace{8mm}

%%%%%%%%%%%%%%%%%%%%%%%%%%%%%%%%%%%
The segmented primary mirror is the defining structural and optical element
of the telescope. A segmented aperture enables a collecting area larger
than that of a monolithic mirror that could be accommodated directly within
the assumed launch-vehicle envelope. Following launch, the individual
segments must operate together as a single optical surface with sufficiently
small piston, tip, tilt, and figure errors to meet the image-quality
requirements of the science instruments.

The primary mirror architecture therefore requires a combination of segment
position sensing, mechanical actuation, optical alignment, and closed-loop
wavefront correction. The alignment system must establish the initial
phasing of the segmented aperture and subsequently compensate for changes
that arise during operation.

KASI has developed and tested segmented-mirror phasing and integration
hardware that provides a useful technical reference for evaluating the
feasibility of the present segmented-mirror concept. 
These activities include the assembly of mirror segments on a
support structure, actuator-based alignment, and interferometric
measurement of segment position and optical phase. The demonstrated
hardware provides a practical starting point for evaluating how these
techniques can be extended to a 3.5-meter-class space telescope.

Future engineering studies would quantify the required segment-position
accuracy, actuator range, sensing precision, and control bandwidth needed
to satisfy the imaging, spectroscopic, and coronagraphic science
requirements.

\section{Optical Performance, Wavefront Control, and Pointing}

The optical system must support several observing modes that place different
demands on image quality and stability. Wide-field imaging requires
well-controlled image quality across an extended field, while
coronagraphic observations require substantially tighter control of
low-order and segment-related wavefront errors. Time-series photometry and
multi-epoch spectroscopy additionally require reproducible instrumental
performance over repeated observations.

Active wavefront control is therefore treated as a telescope-level
function. The control architecture must measure and correct residual
segment misalignment, slowly varying optical errors, and other
wavefront perturbations that affect science performance. 

Pointing control is likewise coupled to optical performance. Spacecraft
attitude control provides coarse and fine target acquisition, while
higher-frequency residual motion must be reduced to levels compatible with
precision imaging and high-contrast observations. Fine-guidance and
optical-path correction methods would be evaluated together with the
spacecraft attitude-control architecture.

Thermal effects are expected to be an important source of slowly varying
wavefront error in a segmented telescope. The structural and optical design
would therefore be assessed for dimensional stability over expected thermal
conditions, including changes associated with spacecraft orientation and
long observing sequences. The relevant thermal requirements would be
derived quantitatively from allowable wavefront drift rather than imposed
as a separate qualitative design principle.

\section{Wide-Field Imaging and Time-Series Capability}

Wide-field imaging provides the principal survey and photometric capability
of the telescope. The current concept targets a field of view between
approximately $10' \times 10'$ and $30' \times 30'$, with the final value
determined by the optical design, detector format, sampling requirement,
and available focal-plane area.

The imaging system is intended to support deep imaging, repeated
photometric monitoring, and high-cadence observations of selected targets.
These observing modes address several parts of the science program,
including galaxy studies, variable and transient sources, exoplanet
photometry, and long-duration stellar monitoring.

Large-format low-noise detectors are being considered for the focal plane.
Detector selection would be driven by quantum efficiency, read noise, dark
current, radiation tolerance, dynamic range, cadence requirements, and
sampling of the telescope point-spread function. Region-of-interest
readout and other subarray modes may be adopted when high temporal
resolution is required.

The final camera architecture would be selected through a trade study
between field of view, detector count, pixel scale, wavelength coverage,
data volume, and photometric performance. Guiding and wavefront-sensing
functions would be defined separately according to the final telescope
control architecture rather than being assumed to reside in a particular
science detector configuration.

\section{Spectroscopic Capability}

Spectroscopy is a core capability of the 3.5-meter Segmented-Mirror Robotic
Space Telescope and is defined primarily by the requirements of its science
programs. The baseline instrument concept covers the optical and
near-infrared wavelength range of 0.2--1.5~$\mu$m and is intended to
complement the wide-field imaging system by providing physical
characterization of sources identified through both survey observations
and targeted programs.

The current planning baseline adopts a spectral resolving power of
approximately $R \sim 1000$ for observations requiring broad wavelength
coverage and high sensitivity. This mode is suitable for galaxy surveys,
classification and temporal evolution of transient sources, and
spectroscopic characterization of relatively faint targets. A higher
spectral-resolution option approaching $R \sim 5000$ is also being
considered for science cases in which resolved line profiles, velocity
structure, or accurate emission- and absorption-line measurements are
important. The final spectral configuration would be determined through
trade studies that balance wavelength coverage, sensitivity, spectral
resolution, detector performance, and instrument complexity. For the
emission-line-galaxy survey, candidate implementations with sufficient
multiplexing or slitless-spectroscopy survey speed would be evaluated
explicitly; the achievable redshift yield would depend on field of view,
target density, exposure time, detector format, and the adopted multiplexing
architecture.

Time-domain spectroscopy places particular requirements on the stability
of the instrumental response between observing epochs. The spectroscopic
system is therefore being developed to control variations in wavelength
solution, spectral line-spread function, detector response, and throughput
over repeated observations. Thermal behavior, opto-mechanical stability,
detector characterization, and wavelength referencing would be evaluated
as part of the integrated instrument design rather than treated as
independent performance parameters.

Calibration requirements are similarly derived from the intended science
measurements. Reference exposures and appropriate internal calibration
sources would be used to track wavelength and detector-response variations,
while observations of astronomical standards can provide complementary
verification of the end-to-end spectrophotometric response. The detailed
calibration architecture remains subject to instrument trade studies, with
the objective of obtaining sufficiently reproducible measurements for
multi-epoch spectroscopy and quantitative comparison of spectra acquired
at different stages of the mission.

The spectroscopic capability supports several distinct components of the
3.5mST science program. For cosmology and galaxy evolution, it provides
redshift measurements and diagnostics of galaxy populations. For
time-domain and multi-messenger astronomy, spectra obtained following
external alerts can identify transient classes and trace their subsequent
evolution. Repeated spectroscopy of white dwarfs, accreting compact
binaries, and other compact-object systems can probe changes in physical
state and velocity structure. Spectroscopic observations can also
complement imaging and high-contrast programs by characterizing selected
planetary and stellar targets.

The spectroscopic instrument concept would continue to evolve together with
the telescope optical design, detector selection, and mission-level science
requirements. The present design therefore specifies the required
observational capabilities and performance ranges while leaving detailed
implementation choices open for subsequent engineering studies.

\section{Coronagraphic Capability}

\begin{figure}[htbp]
  \centering
  \includegraphics[width=0.90\linewidth,height=0.45\textheight,keepaspectratio]{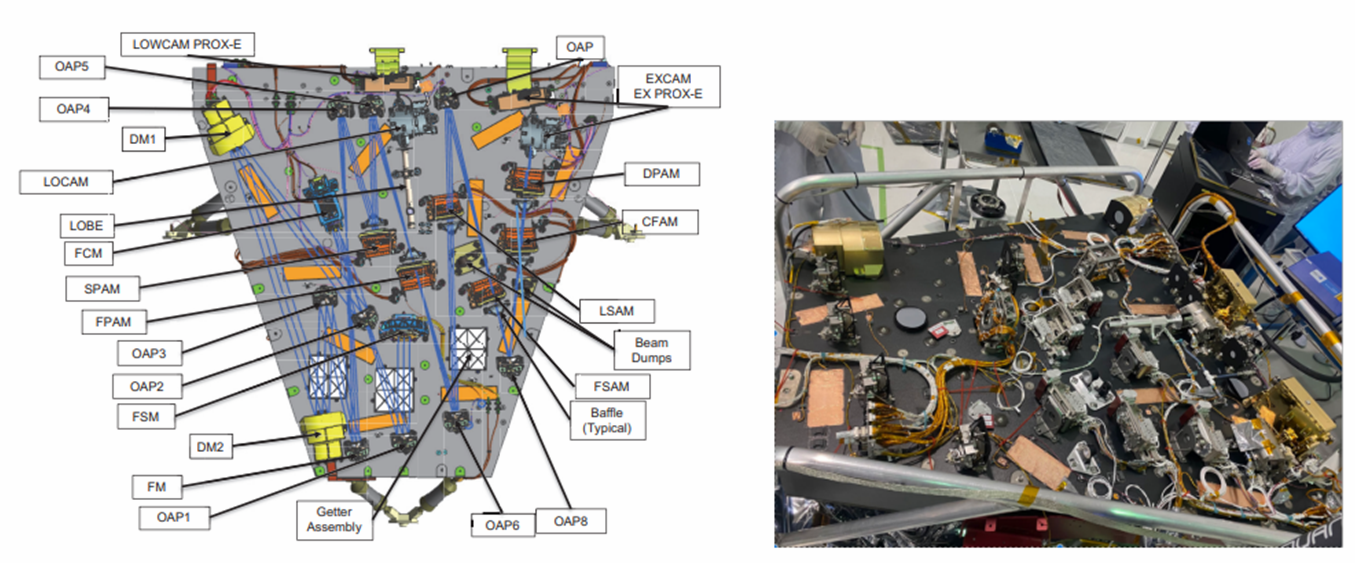}
  \caption{Exoplanet coronagraph of the Nancy Grace Roman Space Telescope
  from Bailey et al. (2023). \cite{bailey2023}}
  \label{fig:roman-coronagraph}
\end{figure}
\vspace{4mm}

Direct imaging of nearby planetary systems is included as one of the major
science capabilities of the telescope. A dedicated stellar coronagraph
is therefore being studied as part of the baseline science payload rather
than as a simple extension of the wide-field imaging instrument.

The present performance goal is a raw star--planet contrast of order
$10^{-8}$. Further improvement through calibration and post-processing is
expected to be investigated as part of the end-to-end coronagraph
performance analysis. The achievable contrast would ultimately depend on
the combined performance of the segmented aperture, wavefront sensing and
control, pointing stability, coronagraph optical design, detector
characteristics, and data-reduction methods.

A segmented primary mirror introduces additional considerations for
high-contrast imaging, including diffraction from segment boundaries and
sensitivity to differential piston and tip--tilt errors. Coronagraph design
studies must therefore be performed using the actual 18-segment pupil
geometry rather than extrapolated directly from monolithic-aperture
systems.

The science program emphasizes nearby planetary systems for which the
angular separation between the host star and planet is most favorable.
Detailed coronagraph studies would determine the required inner working
angle, usable wavelength bands, wavefront stability, and observing cadence
for both planet detection and subsequent characterization.

The coronagraphic program also provides a technology-development pathway
for future high-contrast space missions. In the present white paper,
however, coronagraph performance is evaluated primarily by its ability to
meet the specific exoplanet science requirements of the 3.5mST mission.

\section{Spacecraft, Orbit, and Mission Environment}

The spacecraft must provide the physical and operational environment
required by the telescope and science instruments. Its principal functions
include attitude control, electrical power, thermal regulation,
communications, command and data handling, and maintenance of the selected
mission orbit. Detailed spacecraft implementation has not yet been fixed
and would be developed in parallel with the telescope and instrument
requirements.

Pointing performance would be driven by the requirements of the science
instruments, especially high-precision photometry and coronagraphic
observations. The spacecraft attitude-control system and the telescope
fine-guidance system would therefore be treated as coupled elements of the
overall line-of-sight control budget.

Thermal-control requirements would likewise be derived from allowable
wavefront drift, detector operating conditions, and instrument stability.
The relative roles of passive thermal design, active thermal control, and
operational constraints would depend strongly on the final orbit and
spacecraft configuration.

The final mission orbit has not yet been selected. The current trade space
includes operations near the Sun--Earth L2 region and an Earth-orbit
configuration. The orbit would be selected through subsequent mission
studies that consider thermal stability, sky accessibility, communications,
servicing options, operational constraints, and overall mission cost.

The communications and onboard data system would be sized according to the
expected data rates of the imaging and spectroscopic instruments, including
high-cadence observing modes. Ground-system architecture, data processing,
and science-operations procedures would be developed after the telescope
and instrument observing requirements are more fully defined.

\section{Development Priorities}

The present architecture should be regarded as a science-driven baseline
rather than a completed engineering design. Several technical questions
must be resolved before the telescope configuration can be frozen.

The highest-priority studies include the demonstration of segmented-mirror
phasing accuracy at the required optical performance, integrated wavefront
and pointing-control simulations, optical design optimization over the
required field of view, detector and focal-plane selection, spectrograph
configuration studies, and end-to-end coronagraph performance analysis.

Additional mission studies are required to compare candidate orbital
configurations, establish thermal and structural stability budgets, define
communications and data-volume requirements, and determine launch and
servicing constraints.

These studies would progressively translate the scientific objectives of
the mission into quantitative subsystem requirements and would determine
which elements of the present baseline should be retained, modified, or
deferred as the 3.5mST concept advances toward a more detailed design phase.

%%%%%%%%%%%%%%%%%%%%%%%%%%%%%%%%%

\chapter{Scientific Objectives}
\section{Cosmology}\label{cosmology}

\subsection{Cosmic Acceleration, Dark Energy, and Dark Matter}\label{cosmic-acceleration-dark-energy-and-dark-matter}

The accelerated expansion of the Universe was first discovered in 1998 through observations of Type Ia supernovae and has since been strongly confirmed by a wide range of cosmological observations \cite{riess98,perlmutter99}. This discovery became a foundation of the standard $\Lambda$CDM cosmological model and led to the interpretation that dark energy accounts for approximately 70\% of the total energy density of the Universe. However, the physical nature of dark energy---whether it is truly the cosmological constant $\Lambda$ or a dynamical field that evolves with time---remains one of the central questions at the frontier of modern cosmology.

The 3.5-meter-class multiwavelength space telescope is designed to address this problem in a manner distinct from all-sky survey missions that maximize statistical coverage. Instead, it would reduce systematic uncertainties in cosmological parameters and enhance physical interpretability through high-resolution, high-precision deep observations. Rather than performing shallow observations over the widest possible area, the telescope would repeatedly survey homogeneous deep fields covering several hundred square degrees, enabling precision tests of both the properties of dark energy and the validity of gravitational theory.

The key observables describing cosmic acceleration include the distance--redshift relation and the cosmic expansion rate, $H(z)$. Type Ia supernovae directly constrain the luminosity-distance--redshift relation through the Hubble diagram, while complementary probes such as BAO and RSD provide additional constraints on $H(z)$ and the growth of structure. By simultaneously utilizing a broad wavelength range from 0.2 to 1.5 $\mu$m, the telescope would enable supernova color corrections and evolutionary corrections to be performed within a single, internally consistent framework. This capability removes major limitations of ground-based observations, including variations in atmospheric transmission and wavelength-dependent calibration uncertainties, and enables stable photometric measurements with a precision better than 0.01~mag.

These measurements would improve constraints on the cosmic expansion history and the dark-energy equation of state when combined with complementary cosmological datasets.
Beyond improving statistical precision, such a dataset would provide a practical means of testing the temporal evolution of dark energy by directly controlling systematic uncertainties associated with color correction and dust extinction.

Another independent probe of cosmic acceleration is provided by Baryon Acoustic Oscillations (BAOs). Acoustic waves that propagated through the primordial plasma of the early Universe have left an imprint on the large-scale distribution of galaxies as a characteristic length scale of approximately 150 Mpc. This standard ruler enables simultaneous measurements of the angular diameter distance, $D_A(z)$, and the Hubble parameter, $H(z)$. The 3.5-meter telescope would conduct spectroscopic observations of emission-line galaxies in the redshift range $1\lesssim z\lesssim3$, while lower-redshift emission-line galaxies remain accessible for complementary studies and survey calibration. The flagship BAO and RSD analysis emphasizes this range. The 3.5-meter telescope will conduct these spectroscopic observations at spectral resolutions of $R \approx 1000 \sim 5000$, enabling the construction of photometric--spectroscopic redshift samples containing millions of galaxies. Although the mission is not intended to perform an ultra-wide all-sky survey, its strategy of deep and homogeneous observations over selected fields provides a valuable independent benchmark for cross-validating the statistical results obtained by missions such as Euclid and the Roman Space Telescope. In particular, when combined with Type Ia supernova distance measurements obtained in the same fields, these observations enable direct tests of the consistency of geometric distance indicators, providing a powerful means of distinguishing among competing dark-energy models.

However, it remains unclear whether cosmic acceleration is driven by a new energy component, commonly referred to as dark energy, or reflects departures from the standard description of gravity on cosmological scales. Discriminating among these possibilities requires complementary measurements of both the expansion history of the Universe and the growth of cosmic structure. The 3.5-meter space telescope addresses these questions primarily through spectroscopic galaxy surveys, Type Ia supernova observations, and complementary gravitational-lensing studies.

A central component of the cosmology program is a wide-field spectroscopic survey of emission-line galaxies (ELGs). The reference survey goal is approximately \(10^{6}\)--\(3\times10^{6}\) secure galaxy redshifts over the principal survey tiers, conditional on achieving the multiplexing, survey speed, sensitivity, and calibration performance required by the final spectrograph architecture. These measurements would map the three-dimensional large-scale structure of the Universe and enable measurements of baryon acoustic oscillations (BAO) and redshift-space distortions (RSD). BAO provides a geometric probe of the cosmic expansion history, while RSD measures the growth of structure through the redshift-dependent quantity \(f\sigma_{8}(z)\). Together, these observables provide complementary tests of the standard \(\Lambda\)CDM cosmological framework and possible departures from it.

Type Ia supernova observations provide an independent probe of the expansion history. The mission emphasizes stable spectrophotometric measurements and rest-frame ultraviolet and optical characterization of Type Ia supernovae, particularly over the redshift range in which reliable subclass identification and systematic control can be achieved. Rather than relying on supernova number counts alone, the program is designed to improve the physical characterization of supernova populations and to reduce astrophysical systematics in their use as cosmological distance indicators. Combined with BAO/RSD measurements and complementary external datasets, these observations would improve constraints on cosmic expansion, structure growth, and the properties of dark energy.

Gravitational lensing provides an additional and complementary route to studying the distribution of matter. The mission does not rely on a dedicated wide-area cosmic-shear survey as its primary cosmological observable. Instead, high-resolution imaging and stable observations can support targeted weak-lensing measurements of selected galaxy groups and clusters, as well as detailed studies of strong gravitational-lens systems. These measurements provide independent information on halo masses, internal mass distributions, and environmental effects and can be combined with spectroscopic redshifts and galaxy-population measurements.

The dark-matter program focuses particularly on the small-scale distribution and internal structure of dark-matter halos. Deep imaging and spectroscopy across void, field, group, and cluster environments would enable controlled comparisons of low-mass galaxy and halo populations under a common selection function. Such measurements can test whether dark matter remains consistent with the cold and effectively collisionless paradigm on dwarf-galaxy scales or whether its small-scale behavior favors alternatives involving self-interactions or wave-like dark matter. Strong-lensing systems provide an additional probe of small-scale mass structure through substructure-sensitive observables, while targeted observations of groups and clusters can provide complementary constraints on halo mass distributions.

The strength of the mission therefore lies in the combination of complementary observables rather than in any single cosmological probe. Galaxy redshift surveys provide three-dimensional maps of large-scale structure and measurements of BAO and RSD; Type Ia supernovae provide independent information on cosmic distances and expansion; and targeted gravitational-lensing observations probe the distribution of matter on galaxy and cluster scales. Joint interpretation of these measurements, together with external cosmological datasets, would provide improved tests of the expansion history, structure growth, dark-energy models, and possible departures from standard gravity.

The mission can also contribute complementary information relevant to measurements of the Hubble constant through its supernova and selected gravitational-lens programs. These observations employ physical information and systematic controls that differ from those of many existing cosmological surveys and can therefore contribute to broader efforts to understand discrepancies among current measurements of the cosmic expansion rate.

In summary, the 3.5-meter space telescope is conceived not simply as a means of increasing all-sky survey statistics, but as a precision astrophysical telescope combining wide-field spectroscopy, stable time-domain observations, deep imaging, and targeted gravitational-lensing studies. Its cosmology program is designed to improve measurements of cosmic expansion and structure growth while providing controlled tests of dark-matter physics across different environments. By combining these capabilities with complementary external datasets, the mission would provide an independent and physically informative contribution to studies of dark energy, gravity, and the nature of dark matter.

\subsection{Unveiling the Nature of Dark Matter}\label{unveiling-the-nature-of-dark-matter}

Dark matter represents one of the most fundamental challenges at the intersection of cosmology and particle physics. Unlike ordinary matter, dark matter interacts only weakly, if at all, through electromagnetic forces, making it effectively invisible to direct observation. Nevertheless, its existence has been established with a high degree of confidence through a wide range of astrophysical observations, including galaxy rotation curves, the internal dynamics of galaxy clusters, gravitational lensing phenomena, and anisotropies in the Cosmic Microwave Background (CMB) \cite{roos2010}. These indirect lines of evidence indicate that dark matter constitutes more than five times the mass of ordinary baryonic matter and serves as the gravitational framework upon which galaxies and large-scale cosmic structures are formed.

The standard $\Lambda$CDM (Lambda Cold Dark Matter) model assumes that dark matter consists of cold, collisionless particles. This framework successfully explains the formation of large-scale structure, the observed properties of galaxy rotation curves, and the detailed features of the CMB with remarkable accuracy. Despite its observational success, however, the fundamental particle nature of dark matter remains unknown, and no viable dark matter candidate has yet been conclusively detected \cite{arbey2021}.

Determining the true nature of dark matter requires a comprehensive approach that combines astrophysical observations, particle-physics experiments, and cosmological datasets. Only through the integration of these complementary avenues of investigation can the physical properties of dark matter be constrained and its role in the evolution of the Universe be fully understood.

\subsubsection{Dark Matter Candidates and Theoretical Frameworks}

No known Standard Model particle can account for the observed abundance and
structure-formation properties of non-baryonic cold dark matter. Consequently,
a wide range of candidates beyond the Standard Model has been proposed.
Prominent examples include Weakly Interacting Massive Particles (WIMPs),
axions and axion-like particles, sterile neutrinos, and ultralight bosonic
dark-matter candidates. These candidates span a broad range of particle masses
and interaction strengths and can lead to distinct astrophysical and
cosmological signatures \cite{baudis2024}.

From a theoretical perspective, dark-matter models can also be distinguished
by their production mechanisms and cosmological evolution. Particles that were
initially in thermal equilibrium may acquire their present-day abundance
through freeze-out, with the resulting relic density determined by quantities
such as the particle mass and interaction cross section. Other possibilities
include non-thermal production mechanisms and dark-sector scenarios involving
additional particles or interactions. These alternatives extend the theoretical
landscape beyond the traditional WIMP paradigm and can produce different
signatures in structure formation and astrophysical observations
\cite{baudis2024}.

\subsubsection{Observational Constraints and Experimental Searches}

Although astrophysical observations provide compelling evidence for the existence of dark matter, they do not directly reveal its microscopic physical properties. Galaxy rotation curves indicate that dark matter forms extended halos surrounding galaxies, a prediction that is broadly consistent with the $\Lambda$CDM framework. On larger scales, gravitational lensing and X-ray observations of galaxy clusters enable the reconstruction of their mass distributions, providing quantitative measurements of the spatial distribution of dark matter.

Experimental searches for dark matter can be broadly divided into three categories. First, direct detection experiments seek signals produced when dark matter particles scatter off atomic nuclei in highly sensitive terrestrial detectors. Second, indirect detection searches attempt to observe products of dark matter annihilation or decay, such as gamma rays, electrons, positrons, or other high-energy particles. Third, collider searches investigate signatures of dark matter production in high-energy particle collisions, typically through measurements of missing energy and momentum. While none of these approaches has yet produced definitive evidence for a dark matter particle, experimental sensitivities continue to improve steadily, placing increasingly stringent constraints on theoretical models \cite{billard2022}.

\subsubsection{Cosmological Data and Sensitivity to Structure Formation}

Numerical simulations of cosmic structure formation demonstrate that the properties of large-scale structure are highly sensitive to the free-streaming length of dark matter particles. For example, hot dark matter, characterized by relativistic particle velocities in the early Universe, suppresses the formation of small-scale structures. In contrast, cold dark matter promotes structure formation even on relatively small scales, allowing galaxies and substructures to emerge efficiently. These differences are reflected in observable properties of the cosmic web, including the distribution of galaxies, filamentary structures, and the abundance of dwarf galaxies.

Recent studies have also explored the possibility of weak interactions between dark matter and neutrinos, challenging the traditional assumption that dark matter is entirely non-interacting apart from gravity. Such interactions could leave detectable imprints on cosmic structure formation and the evolution of density fluctuations. In addition, gravitational lensing observations have begun to reveal evidence for small-scale dark matter clumps within and around galaxies, providing new datasets that can be directly compared with the predictions of the $\Lambda$CDM model and its alternatives.

Together, these observational and theoretical developments highlight the growing importance of combining cosmological observations, astrophysical measurements, and particle-physics experiments to uncover the fundamental nature of dark matter and its role in shaping the evolution of the Universe.

\subsubsection{Future Prospects and Cosmological Experiments}

Deep and high-precision cosmological observatories, such as a 3.5-meter space telescope, have the potential to probe the fine structure and dynamical properties of dark matter with unprecedented accuracy. For example, by analyzing weak and strong gravitational lensing distortions, it is possible to reconstruct the three-dimensional structure of dark matter halos and thereby place indirect constraints on the interaction properties predicted by particle physics models. When combined with ground-based experiments, these cosmological approaches would play a crucial role in unveiling the true nature of dark matter.

\subsection{Initial Conditions of the Universe and Their Empirical Tests}\label{initial-conditions-of-the-universe-and-their-empirical-tests}

The initial conditions of the Universe and their empirical verification constitute one of the central challenges in cosmology. Our understanding of the formation and evolution of cosmic structures originates from these primordial conditions. The early Universe existed in an extremely hot and dense state and subsequently evolved through rapid expansion and cooling, eventually giving rise to the large-scale structures observed today. The initial conditions are primarily tested through comparisons between theoretical predictions and observations of the anisotropies in the Cosmic Microwave Background (CMB), the distribution of large-scale structures, and primordial nucleosynthesis. These observational tests form the foundation of the $\Lambda$CDM cosmological model and play a crucial role in characterizing the physics of the early Universe \cite{akrami2020}.

The standard paradigm of modern cosmology assumes that the initial conditions of the Universe originated from density fluctuations described by a random Gaussian field. These fluctuations are believed to have been generated during the inflationary epoch, a process in which microscopic quantum fluctuations were amplified into macroscopic density perturbations spanning cosmological scales. Inflation not only provides a mechanism for generating the primordial fluctuations that seed cosmic structure formation but also offers elegant solutions to the horizon, flatness, and homogeneity problems. Consequently, it has become the leading theoretical framework for explaining the initial conditions imprinted in both the CMB and the large-scale structure of the Universe \cite{baumann2011}. The statistical properties of these primordial fluctuations and their subsequent evolution are preserved as precise patterns in the temperature and polarization anisotropies of the CMB.

CMB observations are often regarded as a direct snapshot of the initial conditions of the Universe. In particular, the \textbf{Planck} satellite provided highly precise measurements of temperature and polarization anisotropies, enabling stringent constraints on the spectrum and statistical properties of primordial fluctuations \cite{planck2020}. The Planck data show excellent agreement with the predictions of the $\Lambda$CDM model, indicating that the primordial perturbations are well described by an almost random Gaussian distribution and exhibit near scale invariance. Specifically, the scalar spectral index, $n_s$, is found to be close to unity, while any detectable level of non-Gaussianity remains extremely small within current observational limits. These results provide important clues for reconstructing the underlying field dynamics that established the initial conditions of the Universe \cite{planck2020}.

Another key avenue for testing the initial conditions is the evolution of large-scale structure. Cosmic structures originated from primordial density fluctuations and subsequently grew under the combined effects of cosmic expansion and gravitational instability, eventually forming the galaxies, galaxy clusters, and vast filamentary structures observed today \cite{peebles1980}. This growth process can be modeled through numerical simulations and compared with observed galaxy distributions, allowing the initial conditions and evolutionary mechanisms to be tested in a unified framework. In particular, the amplitudes and correlations of fluctuations present in the primordial density field are transformed through gravitational interactions and are reflected in the present-day two-point correlation function of matter. Consequently, the statistical properties of the observed matter distribution provide an additional and independent window for testing primordial fluctuations \cite{springel2005,dodelson2003}.

Primordial nucleosynthesis provides another powerful probe of the early Universe by predicting the relative abundances of hydrogen, helium, and lithium. These predictions are highly sensitive to the density, temperature, and baryon content of the Universe during its earliest stages \cite{schramm1998}. Observations of elemental abundances are generally consistent with theoretical expectations, enabling constraints to be placed on the thermodynamic conditions and expansion rate of the early Universe. In particular, primordial nucleosynthesis offers an important and independent determination of the baryon density, complementing constraints derived from CMB observations.

Taken together, CMB anisotropies, the large-scale distribution of matter, and primordial nucleosynthesis provide complementary tests of the initial conditions of the Universe. For example, the CMB directly measures the primordial fluctuation spectrum, whereas large-scale structure observations reveal how that spectrum evolved over cosmic time. By combining these observational probes, cosmologists can determine the initial conditions and cosmological parameters with significantly greater precision than would be possible using any single dataset alone \cite{tegmark2004,komatsu2011}. Such joint analyses enable high-confidence tests of cosmological models that cannot be achieved through individual observational probes in isolation.

Another important prediction of inflationary theory is the possible existence of \textbf{non-Gaussianity} in the primordial fluctuations. While the majority of primordial perturbations are expected to follow a Gaussian distribution, many inflationary models predict small but measurable deviations from Gaussianity \cite{bartolo2004}. Precision analyses of CMB and large-scale structure data place increasingly stringent constraints on such non-Gaussian signatures, providing a powerful means of distinguishing among competing classes of inflationary models \cite{planck2020,akrami2020,meerburg2019}. Measurements of the level of non-Gaussianity therefore offer direct insights into the interactions and dynamics of the fields that governed the early Universe.

The verification of primordial conditions has also benefited significantly from measurements of the polarization of the Cosmic Microwave Background. In particular, \textbf{B-mode polarization} provides a potential direct probe of the primordial gravitational-wave background, which is predicted to have been generated during the earliest stages of inflation \cite{kamionkowski2016}. Although the detection of a primordial B-mode signal remains challenging because of contamination from Galactic dust emission, future CMB polarization experiments and high-precision space-based observatories are expected to improve the separation of foreground signals and enhance the prospects for a definitive detection.

Precision cosmological observatories such as the proposed telescope can further advance studies of the early Universe by directly probing the epoch of reionization and the formation of the first cosmic structures after the CMB era. For example, observations of the distribution of galaxies and galaxy clusters at intermediate to high redshifts, combined with measurements of the neutral hydrogen 21-cm line, can provide a deeper and more comprehensive window into the initial conditions of the Universe and their subsequent evolution \cite{furlanetto2006,pritchard2012}. Such observations extend beyond simply testing the expansion history or the growth of density fluctuations; they form part of a \textbf{multimodal observational strategy} aimed at uncovering the fundamental physics of the early Universe through multiple complementary probes.

\section{Galaxies}\label{galaxies}

\subsection{Cosmic Reionization, Large-Scale Structure, and Cosmological Probes}\label{cosmic-reionization-large-scale-structure-and-cosmological-probes}

The \textbf{Epoch of Reionization (EoR)} refers to the period in cosmic history following the era when the Universe remained largely neutral ($z\gtrsim6$), during which ultraviolet radiation emitted by the first generations of stars and galaxies reionized the hydrogen in the intergalactic medium (IGM). This epoch had profound consequences for the formation of large-scale cosmic structures and the evolution of the free-electron density in the Universe. It also serves as a critical astrophysical background for interpreting cosmological observables such as the matter distribution, gravitational lensing signals, and secondary anisotropies of the Cosmic Microwave Background (CMB). Understanding the processes of reionization and the associated formation of cosmic structures is therefore one of the fundamental challenges of modern cosmology \cite{ahn2019}.

Reionization is the outcome of a highly complex interplay of astrophysical processes. Star formation within galaxies, subsequent supernova explosions, feedback from active galactic nuclei (AGN), and the influence of cosmic rays all affect the escape fraction of ionizing ultraviolet photons and the thermal and dynamical state of the surrounding medium. Recent studies have demonstrated through numerical simulations that internal galactic feedback processes influence not only the evolution of the EoR itself but also the overall distribution of matter in the Universe. For example, galaxy-formation feedback can suppress star formation efficiency, thereby modifying both the timing and the amplitude of reionization. These effects, in turn, alter the interpretation of cosmological observables derived from large-scale surveys and precision cosmological measurements \cite{farcy2025}.

The state of cosmic reionization in the early Universe is reflected in several key cosmological observables. In particular, the post-reionization scattering signature imprinted on the Cosmic Microwave Background (CMB), including its E-mode polarization and temperature anisotropy patterns, is closely linked to the cumulative evolution of the free-electron density over cosmic time. High-redshift objects, such as distant quasars and Lyman-$\alpha$ emitting galaxies, provide valuable observational windows into the inhomogeneous transition from a neutral to an ionized Universe \cite{smith2022}.

Cosmic reionization and the formation of large-scale structure (LSS) are intimately connected. Structures originating from primordial density fluctuations grow through gravitational instability, eventually forming the vast cosmic web of filaments, galaxy clusters, and superclusters observed today. This process not only constrains cosmological parameters---such as the matter density parameter ($\Omega_m$), characteristic clustering scales, and the nature of gravity---but also leaves observable imprints from reionization. In particular, the size and distribution of ionized regions generated during reionization influence various LSS observables, including the matter power spectrum and galaxy environments. Reionization can suppress baryonic clustering on small scales, thereby modifying the evolution of cosmic structures.

More specifically, feedback mechanisms---including active galactic nucleus (AGN) outflows, supernova explosions, and cosmic-ray pressure---can redistribute baryonic matter beyond the boundaries of their host dark matter halos. Compared with models that consider only gravitational evolution, these processes suppress power on small scales in the matter power spectrum. Such suppression can introduce systematic biases when interpreting precision measurements of large-scale structure in combination with CMB observations, including weak gravitational lensing (cosmic shear) as well as the kinetic and thermal Sunyaev--Zel'dovich (kSZ and tSZ) effects \cite{siegel2025}.

Reionization itself is governed not only by internal galactic processes but also by local environmental conditions, such as proximity to massive galaxy clusters and the distribution of collapsed dark matter halos. These factors determine important characteristics of the reionization process, including the sizes of ionized bubbles and the spatial inhomogeneity of ionization fronts. Such inhomogeneities can be directly probed through observations of the Lyman-$\alpha$ forest and the neutral hydrogen 21-cm line, providing a means of simultaneously tracing both cosmic structure formation and the physical mechanisms driving the Epoch of Reionization \cite{smith2022}.

Within this framework, the estimation of cosmological parameters---such as the Hubble constant, the matter and dark energy densities, and the scalar spectral index ($n_s$)---cannot rely solely on observational data. Variations in the matter distribution induced by galaxy feedback processes and reionization can significantly distort cosmological signals. Consequently, modern cosmology increasingly emphasizes the systematic modeling of the interplay between reionization and large-scale structure formation. Hydrodynamical simulations and halo-based modeling have become essential tools in this effort. By jointly accounting for feedback strength, halo mass functions, and the relationship between galaxy mass and environment, these simulations enable quantitative assessments of how reionization influences observable matter distributions and cosmological signals \cite{smith2022}.

For example, measurements of the kinetic Sunyaev--Zel'dovich (kSZ) effect probe the distribution and motion of free electrons associated with cosmic structures. When combined with gravitational-lensing and X-ray measurements, kSZ observations can constrain baryonic-feedback models that redistribute gas within and beyond halos. These same feedback processes can suppress the matter power spectrum on small scales and therefore constitute an important systematic in precision weak-lensing cosmology \cite{siegel2025}.

Ultimately, a variety of physical processes occurring within galaxies---including supernova feedback, AGN-driven winds, and ionizing radiation associated with reionization---play central roles in shaping the reionization history of the Universe. The consequences of these processes are imprinted on large-scale structure and a wide range of cosmological observables. As a result, studies of reionization and large-scale structure have evolved beyond a purely astrophysical problem and now constitute an integrated observational and theoretical framework for constraining cosmological parameters and testing the initial conditions of the Universe.

\section{Compact Objects}

\subsection{Investigating White Dwarf Internal Matter Properties with a 3.5 m Space Telescope}

White dwarfs are the dense stellar remnants left behind after stars with masses comparable to that of the Sun complete their evolutionary lifecycles. Their interiors are supported by electron degeneracy pressure, representing an extreme state of matter that cannot be reproduced under terrestrial laboratory conditions. These environments contain highly compressed plasma and provide a unique laboratory for studying fundamental physics at the intersection of cosmology, particle physics, and astrophysics.

Most white dwarfs possess cores composed primarily of carbon (C) and oxygen (O), although more massive progenitors may produce oxygen--neon (O--Ne) cores in extreme cases. The core composition is closely linked to the nuclear-burning history of the progenitor star. In particular, the carbon-to-oxygen ratio and the detailed distribution of chemical elements depend strongly on the star's initial mass and the rates of nuclear reactions that occurred during its evolution. These properties subsequently influence the cooling timescale of the white dwarf, the progression of crystallization within its interior, and various aspects of its surface and atmospheric physics \cite{salaris1997}.

One of the most powerful tools for investigating the internal properties of white dwarfs is asteroseismology. Pulsating white dwarfs exhibit non-radial oscillation modes generated within their interiors, and the frequencies of these oscillations are highly sensitive to the internal density structure, chemical stratification, and temperature distribution. By comparing observed pulsation periods and mode characteristics with theoretical models, it is possible to place quantitative constraints on core composition, the state of degenerate electrons, thermal conductivity, and other fundamental physical properties of the stellar interior. For example, the comprehensive review by Córsico (2019) \cite{corsico2019} describes how different pulsation modes effectively ``scan'' the deep interior of a white dwarf, providing estimates of quantities such as the stellar mass, rotation rate, and magnetic-field strength.

Studies of white dwarf interiors are closely connected to their core composition and cooling history. After nuclear fusion ceases, a white dwarf gradually cools by radiating away its residual thermal energy over timescales of billions of years. The cooling process is governed by the interplay between conductive energy transport through degenerate electrons and radiative energy loss from the stellar surface. Temperature gradients between the core and the surface, together with diffusion and convection at compositional boundaries, influence both the cooling rate and the observed pulsation modes. Detailed analyses of pulsation signals therefore make it possible to constrain the equation of state of matter in the stellar interior---information that cannot be obtained through measurements of luminosity and temperature alone \cite{bedard2024}.

A 3.5 m space telescope would provide significant advantages for studying the internal properties of white dwarfs in at least three major ways.

First, it would enable high-precision analyses of pulsation modes through photometric and spectroscopic observations. The pulsation periods of white dwarfs, typically on the order of several hundred seconds, are closely linked to atmospheric properties, core composition, and the state of electron degeneracy pressure. Equipped with highly sensitive instruments, a space telescope can obtain long-duration, high-duty-cycle light curves free from atmospheric seeing and transmission variations, allowing the phases and amplitudes of pulsation modes to be measured with high precision. Space-based time-series missions such as Kepler/K2 and TESS have already transformed white-dwarf asteroseismology. A next-generation space telescope would be capable of detecting subtler oscillation signals, thereby enabling increasingly detailed constraints on the internal structure of white dwarfs \cite{corsico2022}.

Second, a 3.5 m space telescope would enable detailed studies of compositional stratification in white dwarfs. Asteroseismic measurements can constrain the core C/O profile and the locations of internal chemical transition zones, while spectroscopy provides complementary information on atmospheric composition. In polluted white dwarfs, detected heavy elements primarily trace accreted planetary debris rather than the composition of the white-dwarf core itself.
Most observed white dwarfs possess atmospheres dominated by either hydrogen (H) or helium (He), while some exhibit atmospheric metal pollution resulting from the accretion of external planetary material. Because heavy elements sink out of the observable atmosphere on element-dependent diffusion timescales, the measured atmospheric abundances do not directly represent the composition of the accreted material. By modeling the accretion history and diffusion of individual elements, however, the bulk composition of the disrupted planetary debris can be inferred, providing valuable information on the composition and differentiation history of extrasolar planetary bodies.

Third, the telescope would provide a unique opportunity to test crystallization processes and the equation of state of degenerate matter. White dwarf interiors reach extremely high densities, typically on the order of $10^6 \sim 10^7 g/cm^3$. Under such conditions, matter can undergo a phase transition from a plasma state to a crystalline state on a macroscopic scale. This crystallization process releases latent heat and produces a measurable cooling delay, an effect that has already been observed in some high-mass white dwarfs. The energy released during crystallization leaves characteristic signatures in the white dwarf luminosity function. Through long-term monitoring and statistical sampling, a space telescope can quantitatively measure these signatures and provide direct empirical tests of dense-matter physics.

The study of white dwarf interior properties also has important cosmological implications. For example, the use of Type Ia supernovae as standard candles is closely linked to the pre-explosion structure of white dwarf progenitors. Internal composition, chemical stratification, and the degree of electron degeneracy can influence the luminosity and color corrections applied to Type Ia supernova observations, thereby contributing to systematic uncertainties in cosmological distance measurements. Consequently, precise studies of white dwarf interiors improve not only our understanding of white dwarf physics but also the accuracy of cosmological parameter determinations derived from Type Ia supernova surveys.

In conclusion, a 3.5 m space telescope would make a unique and transformative contribution to the study of white dwarf interior matter. Its ability to perform long-duration, high-duty-cycle observations free from atmospheric interference, combined with high-sensitivity spectroscopic capabilities and broad wavelength coverage, would significantly strengthen observational constraints on compositional stratification, the equation of state of degenerate matter, and the cooling and crystallization processes occurring within white dwarfs. These data would enhance our understanding of white dwarf physics while also supporting a wide range of cosmological applications, including the calibration of Type Ia supernova standard candles and the reconstruction of the star-formation and evolutionary history of galaxies.

\subsection{Analyzing Circumstellar Material Around High-Density Compact Objects}\label{analyzing-circumstellar-material-around-high-density-compact-objects}

\begin{figure}[htbp]
  \centering
  \includegraphics[width=0.75\linewidth,height=0.75\textheight,keepaspectratio]{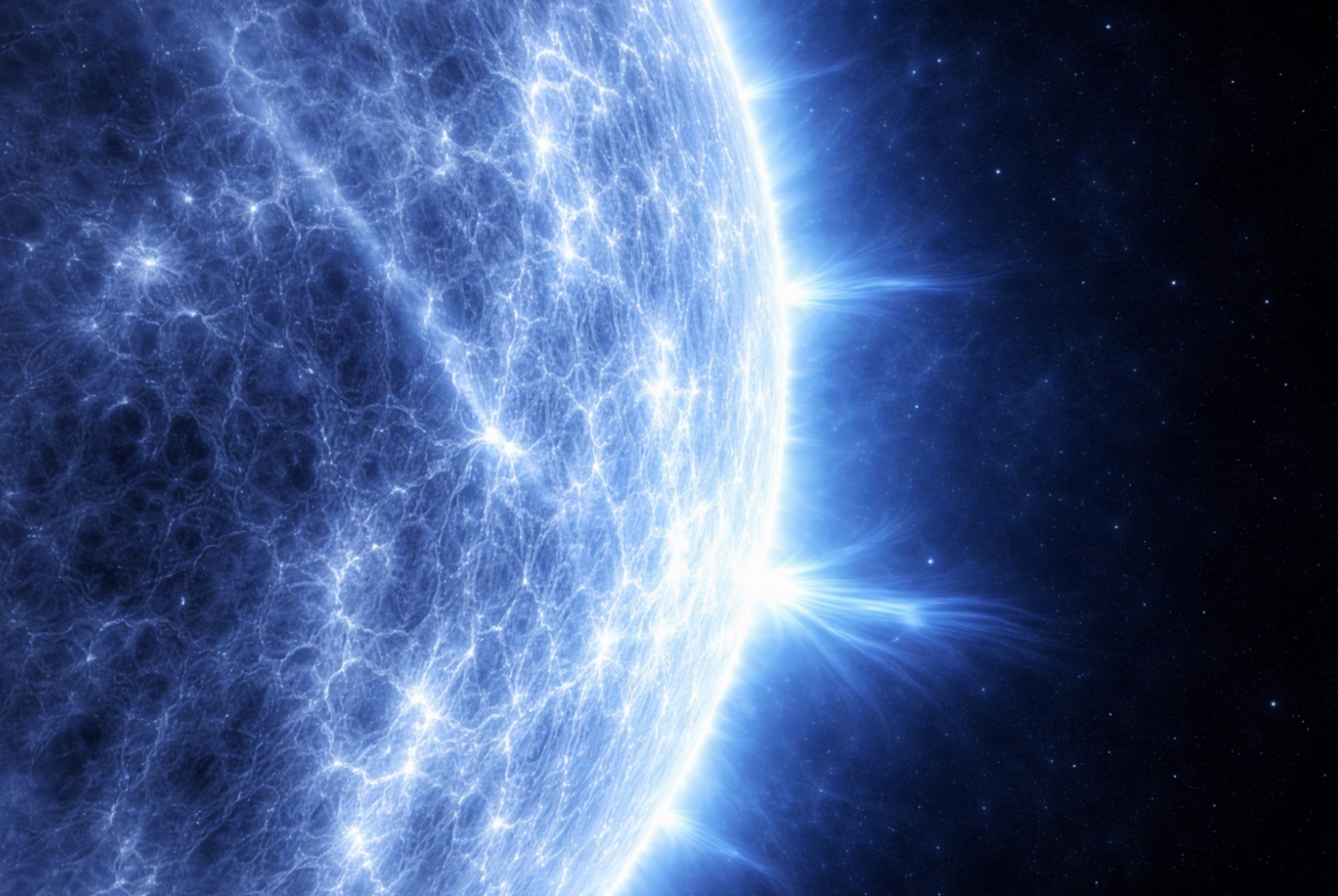}
  \caption{AI-generated conceptual illustration of a white dwarf.}
  \label{fig:white-dwarf-concept}
\end{figure}

The distribution and physical properties of circumstellar material around high-density compact objects, particularly white dwarfs, preserve valuable records of planetary systems that survived the final stages of stellar evolution. Because white dwarfs possess extremely strong gravitational fields, the detection of metal accretion onto their surfaces indicates the presence of material that remains in stable orbit around the star. This material is generally believed to originate from remnant planetary systems, such as asteroid belts, minor planets, or disrupted planetary bodies that survived the host star's evolution \cite{koester2014,debes2012}.

A primary driver of circumstellar material around white dwarfs is planetary debris. Following stellar evolution, surviving asteroids or planetary remnants may be perturbed onto highly eccentric orbits that bring them close to the white dwarf's Roche limit. Once inside this critical distance, tidal forces can disrupt these bodies, fragmenting them into smaller particles and gas. The resulting debris is redistributed into a circumstellar debris disk surrounding the central star \cite{debes2012}. Such disks are frequently detected through infrared excess emission and provide strong evidence for the long-term presence of circumstellar material around white dwarfs \cite{zuckerman1987}.

The detection of heavy elements such as calcium (Ca), magnesium (Mg), and iron (Fe) in white dwarf atmospheres is commonly referred to as metal pollution\textbf{.} This phenomenon is particularly significant because the strong gravitational fields of white dwarfs should cause heavy elements to settle out of the atmosphere on relatively short timescales. The continued presence of these metals therefore implies ongoing replenishment by external material \cite{koester2014,farihi2010}. In other words, material from the surrounding debris disk is continuously accreted onto the white dwarf surface. Measurements of the atmospheric abundances of these elements provide a unique opportunity to infer the composition and distribution of the disrupted planetary material from which they originated.

Collisions and tidal disruption are considered the principal mechanisms responsible for disk formation and material redistribution. For example, observations of the system WD 1145+017 have revealed the ongoing disintegration of small planetary bodies in close proximity to a white dwarf \cite{vanderburg2015}. These events are observed as irregular transit dimming signatures caused by clouds of dust and debris passing in front of the star. In addition, interactions among dust particles and gaseous material within the disk play a critical role in determining the dynamical evolution of the system. Several systems observed with the Spitzer Space Telescope have exhibited variations in disk brightness and color that are associated with changes in particle-size distributions, demonstrating the importance of disk dynamics and matter interactions within circumstellar debris environments \cite{xu2018}.

\subsubsection{Material Distribution and Composition Analysis}

The chemical composition of circumstellar material around white dwarfs provides important clues about the nature of its parent bodies. A study by Xu et al. \cite{xu2019}, based on an analysis of 19 white dwarf systems, found that the chemical abundances of the surrounding material exhibit patterns similar to those of terrestrial planets in the Solar System, with relatively consistent ratios of iron (Fe), magnesium (Mg), and calcium (Ca). These results suggest that the material surrounding white dwarfs is predominantly composed of rocky, terrestrial-like matter. Such analyses are derived from high-resolution spectroscopic observations and are inferred from the relative abundances of metallic elements detected in white dwarf atmospheres. By comparing these atmospheric abundances with models of diffusion and accretion, it is possible to reconstruct the composition of the disrupted planetary material from which they originated.

\subsubsection{Mass Accretion and Condensation Processes}

Theoretical studies of the mechanisms by which circumstellar material is transported onto white dwarf surfaces are closely linked to the Poynting--Robertson (PR) drag effect. Rafikov \cite{rafikov2011} demonstrated through numerical modeling that PR drag causes disk particles to gradually lose angular momentum and spiral inward toward the white dwarf. During this inward migration, dust particles may undergo sublimation, transforming into gaseous material that can ultimately accrete onto the stellar surface. This process provides a natural pathway for the continuous replenishment of metals observed in polluted white dwarf atmospheres.

Subsequent studies have expanded upon this framework by examining the coupled evolution of gas and dust within circumstellar disks. These investigations have proposed mechanisms involving gas--particle interactions, enhanced inward acceleration, and runaway accretion, which can lead to unstable mass-transfer episodes and significantly elevated accretion rates \cite{rafikov2011}. Such processes are believed to play a key role in shaping the dynamical evolution of debris disks and in regulating the flow of planetary material onto white dwarfs. Understanding these accretion mechanisms is therefore essential for interpreting the observed properties of polluted white dwarfs and for reconstructing the long-term evolution of remnant planetary systems.

\subsubsection{Disk Structure and Overall Dynamics}

Observationally, circumstellar disks around white dwarfs consist of gaseous material and dusty debris and are most commonly detected through their infrared excess (IR excess) emission. Well-known examples include the white dwarfs G29-38 and GD 362. In particular, infrared observations of G29-38 with the Spitzer Space Telescope revealed that a significant fraction of the dust particles is concentrated near the white dwarf's Roche radius. This system has become one of the canonical examples for understanding the processes of disk formation and the redistribution of tidally disrupted planetary material into a circumstellar disk.

Furthermore, the observed correlation between the degree of metal pollution and the presence of circumstellar disks provides quantitative information on accretion rates and the distribution of surrounding material. Interestingly, some white dwarfs exhibit clear atmospheric metal pollution despite showing no detectable debris disk. This suggests the existence of tenuous or low-mass debris populations that remain below current observational detection limits \cite{rocchetto2015}. Such cases highlight the continuing importance of improving observational sensitivity and refining models of particle-size distributions in studies of circumstellar material around white dwarfs.

\subsubsection{Significance of Observations with a Space Telescope}

A 3.5 m-class space telescope, with stable space-based photometric and spectroscopic capabilities over a broad wavelength range, would provide a powerful platform for investigating the composition, structure, and temporal evolution of circumstellar material around white dwarfs. In particular, repeated spectroscopy and multi-band monitoring could trace metal absorption features, changes in accretion signatures, and variability associated with debris disks and disrupted planetary material across a substantially larger sample of systems.

Such observations would complement infrared studies obtained with facilities such as the James Webb Space Telescope and the Spitzer Space Telescope by providing systematic, long-term monitoring and spectroscopic characterization from the ultraviolet through the optical and near-infrared. By combining measurements of atmospheric metal pollution, accretion variability, and circumstellar emission or absorption, the mission could constrain the composition of disrupted planetary bodies and investigate how remnant planetary systems evolve after their host stars become white dwarfs.

\subsubsection{Summary}

In summary, the study of circumstellar material around white dwarfs extends far beyond the simple detection of remnant debris. It represents a crucial field of research for understanding the physical state and evolutionary mechanisms of matter surrounding high-density compact objects from both astrophysical and cosmological perspectives. These investigations provide essential constraints on the final stages of planetary system evolution, the formation and evolution of debris disks, and the relationship between accretion rates and material composition in polluted white dwarf systems. As observational capabilities continue to improve, studies of circumstellar material would play an increasingly important role in revealing the fate of planetary systems and the complex interactions between compact stars and their surrounding environments.

\subsection{Particle Interaction Physics in Dense Compact Objects}\label{particle-interaction-physics-in-dense-compact-objects}

Dense compact objects provide complementary laboratories for high-density matter physics. Neutron-star interiors may contain nucleonic superfluids, hyperons, deconfined quark matter, or color-superconducting phases, whereas white dwarfs primarily probe strongly coupled Coulomb plasmas, electron-degenerate matter, and crystallization.
These properties can be investigated observationally through their effects on the \textbf{equation of state (EOS)} of dense matter, gravitational-wave signatures, magnetic-field evolution, and thermal cooling behavior \cite{lattimer2007,ozel2016}.

Research into particle interactions within dense compact objects extends beyond conventional nucleon--nucleon interactions and encompasses a wide range of possible states of matter, including strange quark matter, deconfined quark matter, potentially including color-superconducting phases and superfluid or superconducting states. These exotic phases directly influence the pressure--density relationship within the stellar interior and consequently affect observable properties such as the mass--radius relation, oscillation modes, and radiative characteristics under extreme physical conditions \cite{baym2018,raaijmakers2020,raaijmakers2021}.

\subsubsection{Equation of State of Dense Matter and Nucleon Interactions}

The equation of state (EOS) of nucleonic matter provides the fundamental starting point for understanding the conditions in the interiors of dense compact objects. In neutron stars, the strong interaction between nucleons is dominated by neutron--neutron (n--n) and neutron--proton (n--p) interactions. These interactions are modeled using modern nuclear physics theories and frameworks derived from quantum chromodynamics (QCD), which describes the behavior of strongly interacting particles at the fundamental level \cite{akmal1998,gandolfi2012}.

Conventional nucleon-based EOS models are considered relatively reliable near the nuclear saturation density, where experimental and theoretical constraints are strongest. However, substantial uncertainties arise at densities exceeding approximately two to three times the nuclear saturation density. In this regime, additional degrees of freedom may become important, including the appearance of hyperons, meson condensates, or deconfined quark matter. In particular, the inclusion of hyperons generally tends to soften the EOS, reducing the maximum mass that a neutron star can support against gravitational collapse \cite{schaffner2008}.

Another possible state of matter predicted by QCD under extreme conditions is the quark--gluon plasma or a phase of color superconductivity. These exotic phases are of particular interest because microscopic particle interactions within them can significantly influence the central pressure and internal structure of neutron stars. Theoretical frameworks such as the MIT Bag Model and the Nambu--Jona-Lasinio (NJL) model are widely used to describe the microscopic interactions and thermodynamic properties of these dense quark-matter phases \cite{alford2008,fukushima2011}.

\subsubsection{Observational Constraints on Particle Interactions from Gravitational Waves}

Since 2015, gravitational-wave observations by the LIGO Scientific Collaboration, Virgo Collaboration, and KAGRA Collaboration have provided powerful new constraints on the equation of state (EOS) and particle interactions in neutron stars through the detection of compact-object merger events. One of the most significant examples is the gravitational-wave event GW170817, which enabled measurements of parameters related to the tidal deformability of neutron stars before and during the merger, thereby placing stringent constraints on the EOS of dense matter \cite{abbott2017,tews2018}.

Tidal deformability quantifies the extent to which a neutron star is distorted by the tidal gravitational field of its companion during a merger. This quantity depends sensitively on the stellar compactness and therefore on the equation of state (EOS) of dense matter. For neutron stars of a given mass, a stiffer EOS generally predicts a larger stellar radius and, consequently, a larger tidal deformability, whereas a softer EOS generally predicts a smaller radius and a smaller tidal deformability. Because gravitational-wave signals are sensitive to these tidal effects, observations of binary neutron star mergers provide an indirect but powerful probe of the microscopic interactions governing dense matter \cite{chatziioannou2020}.

\subsubsection{Neutron Star Cooling and Particle Interactions}

The thermal evolution of neutron stars after their formation provides another important observational window into the physics of dense matter. Newly formed neutron stars cool rapidly during their first few minutes through intense neutrino emission. Subsequently, over timescales of thousands to hundreds of thousands of years, their thermal evolution can be tracked through observations of surface X-ray emission and temperature decline curves. The detailed cooling history depends strongly on microscopic physical processes, including nucleon interactions, superfluid energy gaps, and the possible presence of hyperons or other exotic particles in the stellar core \cite{page2006,yakovlev2004}.

For example, nucleon superfluidity induces pairing effects among neutrons and protons within the stellar interior. These pairing interactions significantly influence neutrino production and emission rates, thereby modifying the cooling behavior of the star. By comparing theoretical cooling models with observational data, it is possible to constrain fundamental parameters such as the critical temperatures associated with superfluid transitions and the strengths of the relevant coupling constants. Such studies provide valuable insights into the microscopic properties of matter under conditions that cannot be reproduced in terrestrial laboratories.

\subsubsection{Constraining Particle--Particle Interactions: Experimental and Theoretical Approaches}

The study of the equation of state (EOS) and particle interactions in dense matter relies on a combination of experimental measurements and theoretical calculations. Experimental investigations of the deconfined quark matter produced in relativistic heavy-ion collision facilities such as Relativistic Heavy Ion Collider and Large Hadron Collider provide valuable insights into the behavior of strongly interacting matter under extreme conditions. Although the thermodynamic regimes explored in these laboratory experiments differ from those found in neutron star interiors, comparisons between the two environments offer important constraints on the properties of dense matter and the underlying strong interaction.

On the theoretical side, a wide range of approaches are employed to model dense matter. Traditional methods based on effective nucleon interaction potentials---including the Skyrme, Gogny, and Relativistic Mean Field (RMF) frameworks---are extensively used alongside approaches derived more directly from quantum chromodynamics (QCD). Among these, the RMF model describes strong interactions through the exchange of meson fields and has become one of the most widely used tools for calculating neutron star mass--radius relations and other macroscopic properties of compact objects. Ongoing comparisons between effective nuclear models and QCD-based approaches play a crucial role in reducing uncertainties in the EOS of dense matter.

\begin{figure}[htbp]
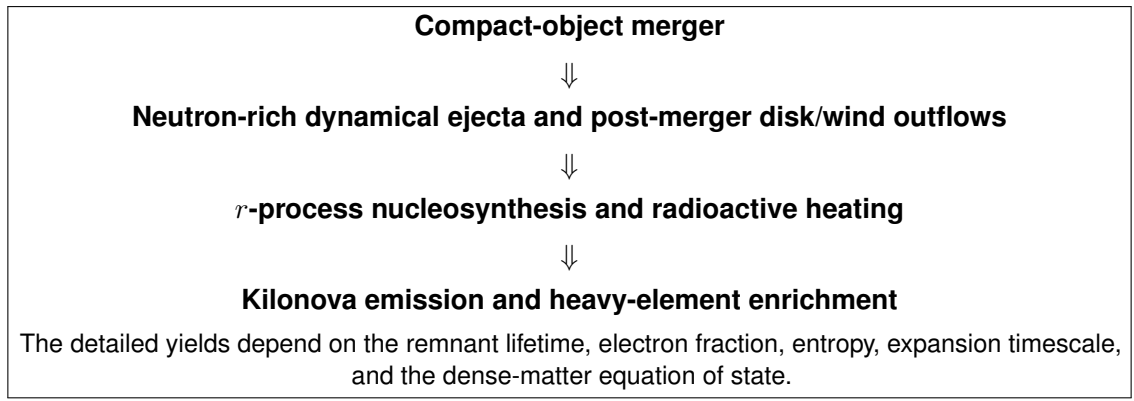

  \centering
  \fbox{\begin{minipage}{0.86\linewidth}
  \centering
  \textbf{Compact-object merger}\\[1.5mm]
  $\Downarrow$\\[1.0mm]
  \textbf{Neutron-rich dynamical ejecta and post-merger disk/wind outflows}\\[1.5mm]
  $\Downarrow$\\[1.0mm]
  \textbf{$r$-process nucleosynthesis and radioactive heating}\\[1.5mm]
  $\Downarrow$\\[1.0mm]
  \textbf{Kilonova emission and heavy-element enrichment}\\[1.5mm]

  \small The detailed yields depend on the remnant lifetime, electron fraction,
  entropy, expansion timescale, and the dense-matter equation of state.
  \end{minipage}}
  \caption{Conceptual schematic of the connection between compact-object mergers,
  neutron-rich ejecta, $r$-process nucleosynthesis, and observable kilonova emission.
  The schematic is provided for illustration and does not reproduce external scientific data.}
  \label{fig:compact-object-nucleosynthesis}
\end{figure}

\subsubsection{Contributions of a 3.5 m Space Telescope}

A 3.5 m-class space telescope could provide new observational constraints on particle interactions in dense compact objects. Its capabilities for high-precision photometry and moderate-resolution spectroscopy, including the optional $R \sim 5000$ mode, would enable sensitive measurements of the surface and spectral properties of neutron stars and white dwarfs. Such observations would provide indirect but powerful tests of microscopic particle interactions and the equation of state governing matter under extreme conditions.

In particular, precision optical timing observations of neutron stars, coordinated with X-ray observations from external space-based facilities, could provide complementary constraints on compact-object variability and internal physical processes. Detailed studies of periodic variability, pulsations, and glitch-related phenomena may help probe internal superfluid components and the interactions of dense matter within neutron stars.

Furthermore, neutron-star mass--radius measurements from X-ray pulse-profile observations and tidal-deformability constraints from gravitational-wave signals provide complementary probes of stellar compactness and the dense-matter EOS. The 3.5mST would contribute primarily through optical/near-infrared timing, spectroscopy, counterpart monitoring, and population studies, which can be combined with X-ray and gravitational-wave measurements from external facilities. Such multimessenger analyses can help discriminate among competing EOS models and improve our understanding of dense matter physics. Achieving these goals therefore requires an integrated strategy combining gravitational waves, X-ray observations, and optical/near-infrared photometric and spectroscopic measurements. Together, these complementary probes offer one of the most promising pathways toward uncovering the fundamental physics of matter at the highest densities found in nature.

\section{Exoplanets}\label{exoplanets}

\subsection{Direct Imaging of Exoplanets Around Nearby Stars Using a Coronagraph}\label{direct-imaging-of-exoplanets-around-nearby-stars-using-a-coronagraph}

Direct imaging is one of the principal techniques for detecting and characterizing exoplanets. It involves suppressing the overwhelming light from a host star in order to isolate and observe the much fainter sources of light originating from orbiting planets. This approach makes it possible to directly visualize exoplanets whose reflected starlight or intrinsic thermal emission is typically millions to billions of times fainter than the light of their parent stars. Unlike traditional indirect detection methods---such as the radial velocity and transit techniques---direct imaging enables direct measurements of a planet's position, brightness, and spectral properties.

The key technologies underlying direct imaging are optical systems designed to suppress stellar light, such as coronagraphs and starshades. A coronagraph blocks or attenuates the stellar image within the optical system, allowing faint nearby objects---including exoplanets and circumstellar disks---to become detectable. These instruments represent one of the most important implementations of high-contrast imaging and are actively developed and employed in both space-based and ground-based observatories \cite{galicher2023}.

\subsubsection{Principles and Limitations of Direct Imaging}

Capturing exoplanets directly around their host stars is one of the most technically challenging methods in observational astronomy. Compared to the Sun, an Earth-like planet is typically more than a billion times fainter, making its detection extremely difficult. In addition, diffraction from the host star and wavefront aberrations within the optical system create significant obstacles to observing such faint signals. A coronagraph mitigates these effects by suppressing diffraction features and reducing scattered starlight, thereby darkening the region surrounding the star in the observer's field of view. This allows the much fainter planetary signal to become relatively more prominent.

The performance of a coronagraph is primarily characterized by two key parameters: contrast and inner working angle (IWA). Contrast represents the ratio between the detectable planetary brightness and the brightness of the host star. Modern coronagraphs and high-contrast imaging systems are being developed to achieve contrast levels as high as approximately $10^{-8}$--$10^{-9}$. The IWA defines the smallest angular separation from the host star at which the coronagraph can effectively suppress starlight. A smaller IWA enables the detection of planets orbiting closer to their stars, including those with smaller orbital radii.

\subsubsection{Current Achievements in Direct Imaging}

Although thousands of exoplanets have been discovered to date, only a relatively small fraction have been detected through direct imaging. Nevertheless, several massive exoplanets---particularly young giant planets such as those in the HR 8799 system (HR 8799 planetary system)---have been successfully observed using this technique. These detections have been made possible through the combination of coronagraphy and advanced adaptive optics systems on large ground-based telescopes.

In 2025, JWST/MIRI coronagraphic observations revealed a sub-Jovian planet candidate around the young star TWA~7  \cite{lagrange25}, with an estimated mass comparable to that of Saturn. Subsequent observations provided additional evidence supporting its planetary nature, and TWA~7~b is now listed as a confirmed planet in the NASA Exoplanet Archive. This object is considered one of the lowest-mass planets ever detected through direct imaging, with a mass comparable to that of Saturn. The result demonstrates that high-contrast imaging technologies are steadily extending their capabilities toward lower-mass planets and targets located at smaller orbital separations from their host stars.

\begin{figure}[htbp]
  \centering
  \includegraphics[width=0.95\linewidth,height=0.95\textheight,keepaspectratio]{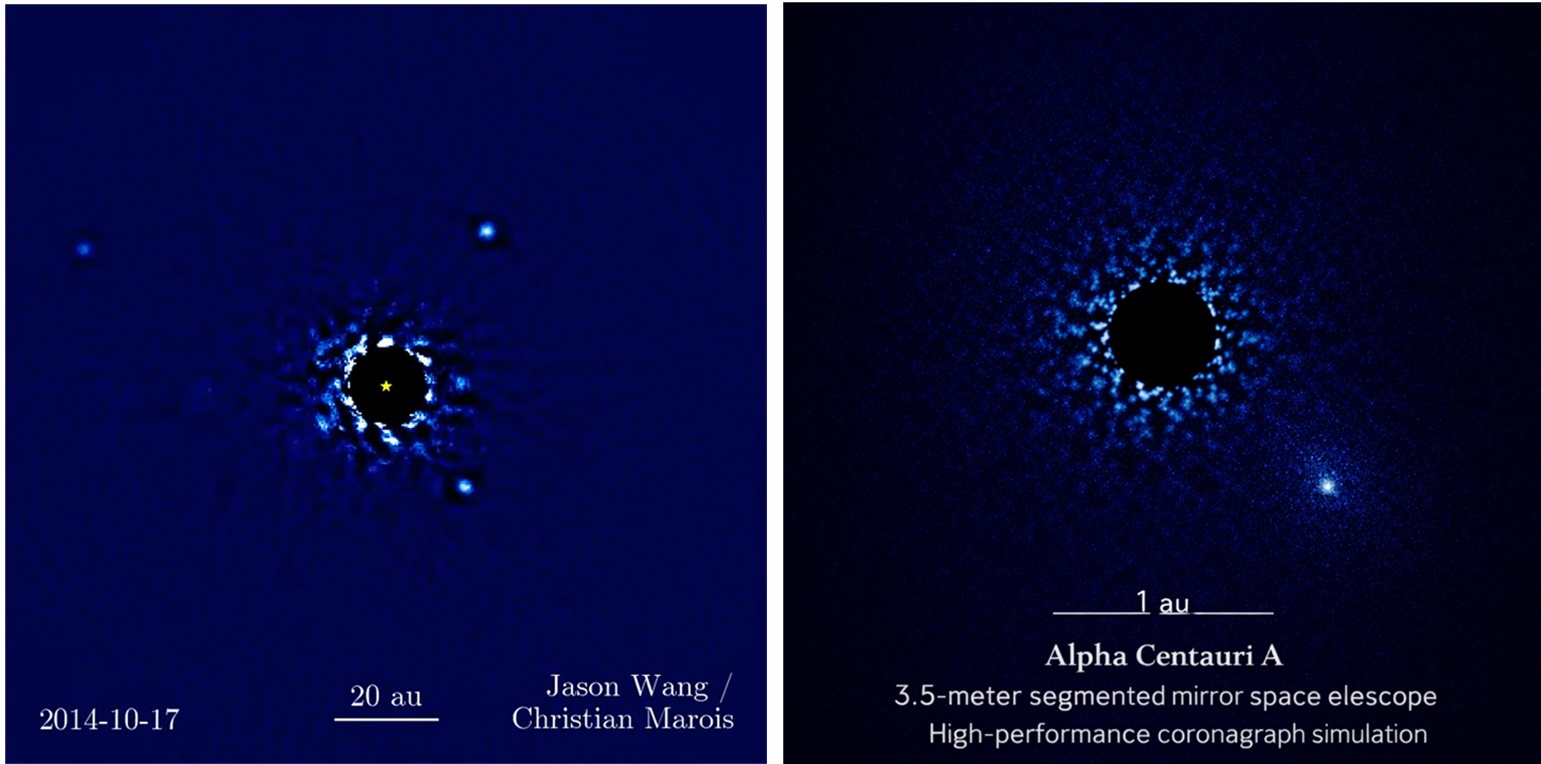}
  \caption{
  Left: a still frame from the HR~8799 orbital-motion visualization
constructed from multi-epoch Keck observations \cite{marois08,marois10,konopacky16};
video credit: Jason Wang and Christian Marois,
with data from W.~M. Keck Observatory; reproduced from
\textit{HR 8799 in Motion (Official)} under the
Creative Commons Attribution 3.0 Unported (CC BY 3.0) license.
Right: an AI-generated conceptual illustration of an exoplanet at
1~AU from Alpha Centauri~A, illustrating a possible appearance in
high-contrast observations with a 3.5-meter segmented-mirror space telescope.
}
  \label{fig:hr8799-alpha-centauri}
\end{figure}

\subsubsection{Direct Imaging Potential of a 3.5-Meter Space Telescope}

A 3.5-meter-class space telescope can expand the applicability of direct imaging by combining diffraction-limited optical performance with a stable, low-background space environment and dedicated stellar-light suppression. A space-based platform is free from atmospheric turbulence and can support long integrations with stable wavefront control, making it well suited to high-contrast observations of nearby planetary systems. When combined with a coronagraph, a 3.5-meter telescope can provide useful capabilities in several key areas.

\begin{itemize}
\item
  \textbf{Detection of Nearby Exoplanets:}\\
  High-contrast imaging can enable the detection of exoplanets at relatively small orbital separations from their host stars, potentially down to a few astronomical units (AU) for nearby systems. This increases the possibility of directly imaging planets in closer-in orbits that are difficult to observe with current direct-imaging facilities.
\item
  \textbf{Spectroscopic Imaging and Atmospheric Characterization:}\\
  By combining a coronagraph with optical and near-infrared spectrographs, it becomes possible to investigate the atmospheric composition of detected exoplanets through the identification of molecular species such as water vapor and methane. Reflected-light spectroscopy can further reveal information about cloud properties, atmospheric structure, and planetary surface conditions. Such observations extend direct imaging beyond simple detection toward comprehensive physical characterization of exoplanets \cite{houlle2021}.
\item
  \textbf{Observations of Planet Formation and Early Evolution:}\\
  Direct imaging is particularly valuable for studying young planetary systems and stellar associations, where newly formed planets remain relatively warm and luminous. Observations at optical and infrared wavelengths can distinguish young giant planets from their host stars, providing insights into planet formation mechanisms and early evolutionary processes.
\end{itemize}

Despite these advantages, significant technical challenges remain for achieving high-performance direct imaging with a 3.5-meter-class space telescope. Wavefront control within the coronagraph system requires extremely precise optical alignment and stability. Achieving deeper contrast levels and reducing the inner working angle to access planets at smaller angular separations demand increasingly sophisticated optical architectures, wavefront-sensing techniques, and post-processing algorithms. These challenges have motivated recent theoretical investigations into advanced approaches, including quantum-information-based signal processing and methods that explore the fundamental quantum limits of high-contrast imaging \cite{deshler2024}.

\vspace{\baselineskip}
\textbf{Future Prospects and Challenges}

Direct imaging is expected to advance along several important directions in the coming decades:

\begin{itemize}
\item
  \textbf{Detection of Small Planets (Earth-like and Super-Earth Planets):}\\
  The baseline coronagraph performance target of the 3.5-meter space telescope is a raw contrast of approximately \(10^{-8}\), with \(10^{-9}\) adopted as an advanced performance goal. Reaching the \(\sim10^{-10}\) contrast regime required for reflected-light observations of Earth analogs is treated as a stretch objective rather than a baseline mission requirement. Progress toward this regime would provide a pathway to future direct detection and characterization of potentially habitable terrestrial planets.  
  Achieving this capability would provide a powerful means of identifying potentially habitable worlds and investigating possible biosignatures.
\item
  \textbf{Atmospheric Spectroscopy and Biosignature Characterization:}\\
  Direct imaging combined with spectroscopic observations can identify molecular absorption features in planetary atmospheres, allowing detailed studies of atmospheric composition and structure. Such measurements may ultimately enable the detection of potential biosignatures and the assessment of planetary habitability.
\item
  \textbf{Multi-Method Observations and Data Integration:}\\
  Direct imaging is most powerful when combined with complementary exoplanet detection techniques, such as transit photometry and radial-velocity (RV) measurements. The integration of these observational methods can provide stronger constraints on key astrophysical parameters, including planetary mass, radius, orbital architecture, atmospheric properties, and evolutionary history.
\end{itemize}

In conclusion, direct imaging with a 3.5-meter-class space telescope has the potential to extend the current technological frontier and enable the detection and characterization of exoplanets around nearby stellar systems. Beyond the discovery of new planets, direct imaging would serve as a powerful scientific tool for investigating planetary atmospheres, physical properties, formation histories, and planetary habitability and the prospects for life. Continued advances in coronagraph technology, wavefront control, and high-contrast imaging systems would further enhance the precision and scientific return of future exoplanet exploration missions.

\subsection{Exoplanet Classification and Population Statistics}\label{exoplanet-classification-and-population-statistics}

In modern astronomy and astrophysics, exoplanet research has evolved beyond the simple discovery of individual planets to the systematic classification of thousands of confirmed exoplanets and the investigation of their statistical properties. The rapidly growing sample reveals a remarkable diversity in planetary masses, radii, orbital characteristics, and system architectures.

The goal of exoplanet classification studies is to organize these planets into meaningful groups based on their physical and orbital properties. Traditionally, exoplanets have been categorized according to their masses and radii into classes such as \textbf{rocky planets}, \textbf{mini-Neptunes}, and \textbf{gas giants}. More recently, classification schemes have also incorporated planetary radius and incident stellar flux to define physically and observationally useful categories for direct-imaging surveys and yield estimates \cite{kopparapu2018}.

In recent years, increasing attention has been directed toward the classification of entire planetary-system architectures. This approach considers not only the physical properties of individual planets but also the relative masses, orbital spacings, and dynamical relationships among multiple planets within the same system. Examples include \textbf{``peas-in-a-pod'' systems}, in which several similarly sized planets occupy compact orbital configurations, and systems containing \textbf{warm Jupiters}. Such classifications provide a more comprehensive description of planetary systems than classifications based solely on planetary mass or orbital period \cite{howe2025}.

Statistical studies focus not only on the characteristics of individual exoplanets but also on determining how frequently different types of planets occur throughout the Galaxy. This field, commonly known as the \textbf{demographics of exoplanets}, seeks to measure occurrence rates and distribution functions as functions of planetary properties (such as mass and radius), orbital parameters (such as period and eccentricity), and host-star characteristics (such as spectral type) \cite{gaudi2020}. These statistical investigations provide critical constraints on theories of planet formation and evolution, helping to reveal the processes that shape planetary systems across a wide range of astrophysical environments.

\subsubsection{Representative Statistical Studies of Exoplanets}

One of the most influential examples of exoplanet population studies is the \textbf{Kepler mission} conducted by NASA. From 2009 to 2013, Kepler continuously monitored approximately 190,000 stars over a period of about four years and discovered thousands of exoplanet candidates. These observations enabled researchers to estimate planetary occurrence rates based on the measured radii and orbital periods of the detected planets. The Kepler dataset demonstrated that small-radius planets are extremely common within orbital distances of roughly 1 astronomical unit (AU) and revealed that planetary systems often exhibit architectures significantly different from that of our Solar System.

Accurate statistical analyses require careful correction for \textbf{selection biases} inherent in exoplanet detection methods. Both the \textbf{transit method} and the \textbf{radial velocity method} possess sensitivities that depend on factors such as planetary size, mass, orbital inclination, and orbital period. Consequently, the observed sample cannot be directly interpreted as representative of the underlying planetary population. To derive reliable occurrence rates, probabilistic correction techniques that account for both \textbf{reliability} and \textbf{completeness} must be applied \cite{bryson2021}.

Exoplanet classification and demographic studies also play a crucial role in testing theories of \textbf{planet formation} and \textbf{planetary system evolution}. By comparing occurrence rates and classification spectra derived from multiple detection techniques with theoretical predictions, researchers can quantitatively evaluate physical processes such as protoplanetary disk mass distributions, star--planet interactions, and planetary migration. The resulting statistical patterns provide insights not only into the abundance of planets but also into the mechanisms that shape planetary systems and their long-term dynamical evolution \cite{gaudi2020}.

In recent years, significant progress has been made in addressing observational biases and improving completeness corrections in exoplanet demographic studies. For example, statistical analyses of the Kepler catalog have incorporated corrections for variations in detection sensitivity and the effects of false positives, enabling the derivation of bias-corrected occurrence rates. These methods have been used to estimate fundamental population parameters, including the occurrence rate of Earth-like planets, commonly denoted as \textbf{$\eta_{\oplus}$} \cite{bryson2021}.

Furthermore, studies of \textbf{planetary-system architectures} have become increasingly important for understanding how multiple planets are arranged and interact within a system. Such investigations examine statistical properties including orbital spacing ratios, orbital resonances, and eccentricity distributions. These measurements help characterize the typical architectures of exoplanetary systems and provide valuable constraints on theories of planetary formation, early dynamical evolution, and large-scale system organization \cite{howe2025}.

\subsubsection{Astrobiological Implications and Future Prospects}

Exoplanet classification and population-statistics studies are also closely linked to \textbf{astrobiology} and investigations of \textbf{planetary habitability}. Large exoplanet datasets provide the fundamental statistical populations required for testing theories of habitability and evaluating how physical and environmental factors influence the potential for life. Such studies are essential for understanding the conditions under which habitable environments may emerge and persist on planetary bodies \cite{hong2023}.

In summary, classification and demographic analyses based on the large samples of exoplanets discovered through both direct imaging and indirect detection techniques provide answers to a broad range of questions concerning planetary diversity, formation and evolutionary processes, and the potential existence of life under specific environmental conditions. These studies extend our understanding beyond individual planetary systems and toward a comprehensive view of planetary populations throughout the Galaxy.

\subsubsection{Expanding the census of exoplanet systems with a 3.5-meter-class space telescope}

Future high-performance observatories, including a \textbf{3.5-meter-class space telescope}, would greatly expand the available exoplanet sample and enable increasingly detailed demographic investigations. Improved sensitivity, higher spatial resolution, and enhanced spectroscopic capabilities would allow the detection and characterization of previously inaccessible planetary populations. Ultimately, these advances would contribute to the construction of a comprehensive map of exoplanetary ecosystems, transforming our understanding from isolated examples to a statistically robust picture of planetary systems across the Universe.

\subsection{Investigating the Habitability of Exoplanets}\label{investigating-the-habitability-of-exoplanets}

The study of \textbf{exoplanet habitability} is one of the most challenging and significant areas of modern astronomy. This field extends beyond merely confirming the existence of exoplanets and seeks to determine whether a planet possesses the physical, chemical, and environmental conditions necessary to support or sustain life. Habitability assessments focus on several key factors, including the potential presence of liquid water on the planetary surface, the existence of stable temperature and pressure conditions, and the detectability of atmospheric \textbf{biosignatures} that may indicate biological activity \cite{robinson2018,schwieterman2018}.

\begin{figure}[htbp]
  \centering
  \includegraphics[width=0.85\linewidth,height=0.42\textheight,keepaspectratio]{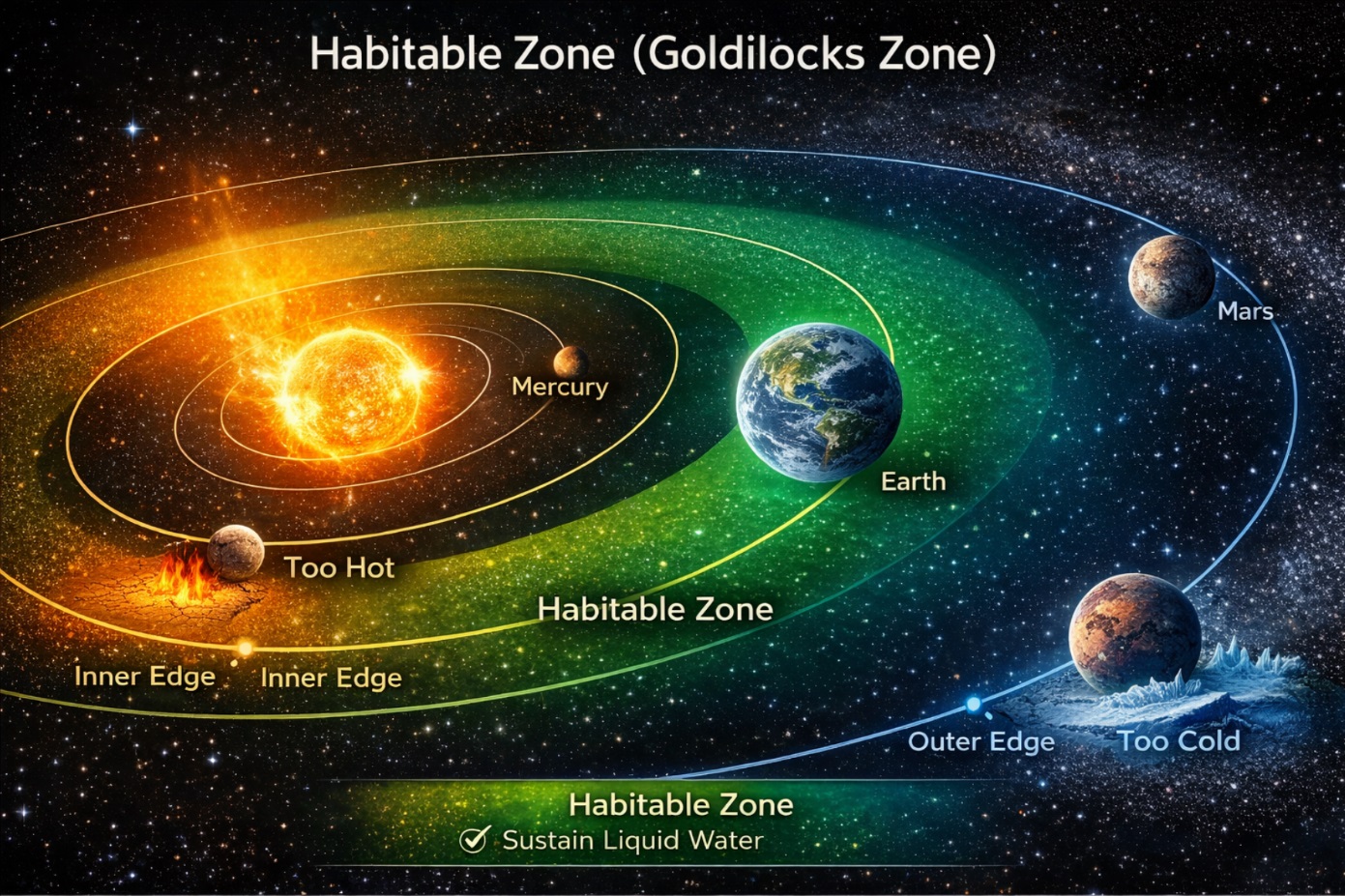}
  \caption{AI-generated conceptual illustration of the habitable zone in the Solar System.}
  \label{fig:habitable-zone-concept}
\end{figure}

\subsubsection{Habitability and the Habitable Zone}

The first step in evaluating planetary habitability is to identify the \textbf{habitable zone (HZ)} around a star. The habitable zone is defined as the range of orbital distances where stellar radiation allows liquid water to exist on a planet's surface. The location and extent of the HZ depend on the spectral type, luminosity, and evolutionary stage of the host star; consequently, the HZ varies with both stellar type and age.

Planets located within the habitable zone are generally considered more likely to satisfy the basic requirement for the presence of liquid water. However, residing within the HZ alone does not guarantee habitability. Additional factors---including atmospheric composition, atmospheric retention capability, greenhouse effects, planetary geology, and the existence of a protective magnetosphere---must also be taken into account when assessing a planet's potential to support life \cite{seager2025}.

When a planet experiences temperature conditions similar to those of Earth, liquid water may remain stable on its surface. Because of water's unique chemical properties and its role as the primary medium for known biological processes, the presence of liquid water is regarded as a fundamental requirement in most habitability studies. Nevertheless, the precise boundaries of the habitable zone are influenced by a variety of factors, including stellar ultraviolet radiation, atmospheric composition, cloud coverage, greenhouse processes, and planetary albedo. These effects are quantified through theoretical and numerical models that refine estimates of habitable-zone limits for different planetary and stellar environments \cite{seager2013}.

\subsubsection{Observational Approaches: Atmospheric Spectroscopy and Biosignatures}

The analysis of exoplanetary atmospheres is one of the most powerful tools for investigating planetary habitability. Atmospheric species such as \textbf{oxygen (O\textsubscript{2}), ozone (O\textsubscript{3}), methane (CH\textsubscript{4}), carbon dioxide (CO\textsubscript{2}), and water vapor (H\textsubscript{2}O)} provide complementary information about possible biosignatures and the environmental context in which they must be interpreted. O\textsubscript{2} and O\textsubscript{3} are widely discussed biosignature candidates, whereas CO\textsubscript{2} and H\textsubscript{2}O are important indicators of atmospheric state and potential habitability. In particular, the simultaneous presence of oxygen and methane is of special interest because these gases tend to react with one another and can be difficult to maintain together in significant quantities without sustained sources. Their coexistence may therefore indicate a state of \textbf{chemical disequilibrium}, although abiotic pathways must be evaluated before any biological interpretation is made.

Such potential indicators of life, known as \textbf{biosignatures}, can be detected through spectroscopic observations performed by space-based telescopes. Both \textbf{transit spectroscopy} and \textbf{direct-imaging spectroscopy} are valuable techniques for identifying biosignature gases within exoplanetary atmospheres \cite{robinson2018,quanz2022}. In transit spectroscopy, starlight passes through the atmosphere of a planet during a transit event, and specific wavelengths are absorbed by atmospheric constituents. By measuring these absorption features, astronomers can infer the chemical composition and physical properties of the atmosphere. This technique has already been successfully applied to characterize the atmospheres of several giant exoplanets using facilities such as the James Webb Space Telescope.

High-contrast imaging combined with \textbf{reflected-light spectroscopy} also plays a crucial role in habitability studies. The reflected-light spectrum of a directly imaged exoplanet contains information about both its atmosphere and surface properties. Such observations can reveal the presence of water vapor, clouds, and a variety of molecular species that may be associated with habitable environments or biological activity. Because reflected-light spectroscopy can probe planetary environments in ways that complement transit observations, it is considered one of the key technologies for future large space-telescope missions dedicated to the search for habitable worlds and extraterrestrial life \cite{damiano2023}.

\subsubsection{Interpretation and Limitations of Biosignatures}

One of the most frequently discussed biosignature candidates is \textbf{molecular oxygen (O\textsubscript{2})}, which constitutes approximately 21\% of Earth's atmosphere and is maintained primarily through biological activity. However, oxygen can also be produced through abiotic processes under certain planetary conditions. Consequently, the detection of oxygen alone cannot be regarded as definitive evidence for the existence of life. For this reason, combinations of multiple biosignatures---such as the simultaneous detection of \textbf{O\textsubscript{2} and methane (CH\textsubscript{4})}---are generally considered more compelling indicators of biological activity. Some researchers have further proposed that the presence of a significant \textbf{thermodynamic or chemical disequilibrium} within a planetary atmosphere may itself serve as a biosignature. To investigate such possibilities, methods have been developed to compare the relative abundances of multiple molecular species and evaluate whether their coexistence can be explained through known abiotic processes \cite{claudi2019}.

Habitability studies must also carefully address the challenges of \textbf{false positives} and \textbf{false negatives}. For example, atmospheric chemical disequilibrium may arise from geological or photochemical processes rather than biological activity, leading to a false-positive interpretation of a biosignature. Conversely, life may exist on a planet without producing atmospheric signatures that are readily detectable, resulting in a false-negative outcome. Therefore, the interpretation of potential biosignatures requires the integration of multiple observational indicators and a comprehensive understanding of the planetary environment in which those signals are observed \cite{schwieterman2018}.

\subsubsection{Contributions and Prospects of a 3.5-Meter Space Telescope}
A 3.5-meter-class space telescope could contribute to habitability studies through high-contrast direct imaging and reflected-light spectroscopy of nearby planetary systems. Its primary exoplanet program would focus on targets accessible within the demonstrated coronagraph performance, while observations of terrestrial planets in nearby habitable zones would remain a challenging objective requiring substantially deeper contrast and sufficient spectroscopic sensitivity.

If the required contrast and spectroscopic sensitivity can be achieved for favorable nearby terrestrial-planet targets, reflected-light spectroscopy could be used to search for atmospheric features associated with molecules such as O$_2$, H$_2$O, and CH$_4$. Such observations should therefore be regarded as an advanced or stretch science objective rather than a guaranteed capability of the baseline mission.

For sufficiently bright and accessible planetary targets, repeated observations may also be used to investigate temporal variations in reflected light and atmospheric properties. The feasibility and scientific scope of such monitoring would depend on the achieved contrast, photometric stability, and available observing time.
These measurements could provide additional clues regarding environmental stability and potential ecological analogs to processes observed on Earth, thereby placing stronger constraints on planetary habitability \cite{robinson2018}.

In summary, the study of exoplanet habitability encompasses multiple interconnected components, including habitable-zone assessment, atmospheric characterization, biosignature detection, and the interpretation of observational evidence. A 3.5-meter-class space telescope represents a powerful platform for advancing all of these areas. Such investigations would move beyond the simple question of whether life exists elsewhere and toward a more comprehensive understanding of the environmental conditions and potential biological processes that may operate on planets throughout the Universe.

\subsection{Stellar Magnetic Activity and Exoplanet Habitability}\label{stellar-magnetic-activity-and-exoplanet-habitability}

\textbf{Stellar magnetic activity} encompasses a wide range of phenomena associated with stellar magnetic fields, including \textbf{starspots}, \textbf{stellar flares}, \textbf{coronal mass ejections (CMEs)}, and \textbf{high-energy X-ray and ultraviolet (XUV) radiation}. These processes can significantly influence the atmospheric composition, atmospheric retention, and surface environment of exoplanets. Their effects are particularly important in planetary systems around \textbf{M-type red dwarfs}, where the \textbf{habitable zone (HZ)} is located very close to the host star. Under such conditions, stellar magnetic activity becomes a key factor in determining planetary habitability \cite{gallet2017,ridgway2023}. Furthermore, investigating the effects of magnetic activity in young solar-type stars provides valuable insights into the evolution of Earth-like atmospheres and the environmental conditions that may have contributed to the origin of life on Earth \cite{namekata2026}.

\subsubsection{Effects of Stellar Magnetic Activity on Planetary Atmospheres}

One of the most prominent manifestations of stellar magnetic activity is the \textbf{stellar flare}, which releases large amounts of high-energy photons and energetic particles. These events can directly affect both the chemical structure and long-term stability of planetary atmospheres. X-rays and extreme ultraviolet (EUV) radiation emitted during flares heat the upper layers of planetary atmospheres, thereby enhancing atmospheric escape processes. If such events occur repeatedly over long timescales, the cumulative loss of atmospheric material can gradually reduce atmospheric thickness and deplete water vapor and other gases essential for maintaining habitable conditions, ultimately diminishing planetary habitability \cite{ridgway2023}.

This effect is particularly severe for planets orbiting M-dwarf stars. Because the habitable zones of these stars are located at small orbital distances, planets within them are exposed to significantly higher levels of X-ray and EUV radiation, making atmospheric retention more difficult. Previous studies have shown that energetic flares from M-type stars can alter or even destroy protective atmospheric layers, such as ozone, leading to substantial long-term changes in atmospheric chemistry \cite{chen2021}.

On the other hand, stellar magnetic activity may also have beneficial effects for the emergence of life. Previous studies suggest that energetic particles associated with stellar activity can promote \textbf{nitrogen fixation}, a process considered important for prebiotic chemistry and the development of early biological systems. Stellar activity may also contribute to the production of greenhouse gases that help maintain favorable surface temperatures \cite{airapetian2016}. In addition, energetic radiation and particle interactions may facilitate chemical pathways leading to the synthesis of amino acids and other prebiotic molecules relevant to the origin of life \cite{kobayashi2023}.

Therefore, the study of stellar magnetic activity is important because it exhibits a dual role in planetary evolution. While intense activity can drive atmospheric loss and reduce habitability, it may simultaneously promote chemical processes that contribute to the emergence of life. Understanding these competing effects not only provides insight into the early evolution of life on Earth but also improves our ability to interpret potential biosignatures and assess the habitability of exoplanets around a wide variety of host stars.

\subsubsection{Stellar Magnetic Activity and the Protective Role of Magnetospheres}

At the planetary scale, a \textbf{magnetosphere} can modify the interaction between an atmosphere and charged particles associated with stellar winds and flare-driven particle events. Depending on the planetary magnetic-field geometry, atmospheric properties, and stellar-wind conditions, magnetic shielding may reduce some forms of direct particle precipitation, while magnetically mediated escape processes can also remain important. Over long timescales, strong stellar-wind and particle forcing may therefore contribute to atmospheric erosion and water loss even for planets located within the habitable zone.

These effects can be particularly important in stellar environments where \textbf{superflares} occur frequently. Consequently, evaluating exoplanet habitability requires a quantitative understanding of the coupled interaction among stellar magnetic activity, the duration and intensity of high-energy events, atmospheric escape processes, and the strength and structure of planetary magnetospheres rather than assuming that a magnetic field always provides complete atmospheric protection.

\subsubsection{Redefining the Habitable Zone in the Context of Stellar Magnetic Activity}

Traditionally, the \textbf{habitable zone (HZ)} has been defined primarily by orbital distance from the host star, based on the assumption that liquid water can exist on a planet's surface within a certain range of stellar irradiation. However, recent studies suggest that the concept of habitability should also incorporate the intensity and temporal variability of stellar magnetic activity.

Under this framework, two planets receiving similar levels of stellar radiation may experience very different environmental conditions if their host stars exhibit different levels of magnetic activity. A highly active star may subject its planets to intense flare activity, elevated XUV radiation, and enhanced stellar-wind interactions, thereby reducing the region in which stable and long-term habitable conditions can be maintained. This concept has led to the proposal of a \textbf{dynamic habitable zone}, in which habitability depends not only on stellar irradiance but also on the surrounding space-weather environment. Such an approach requires consideration of stellar age, rotation rate, magnetic-field strength, and long-term activity evolution as integral components of habitability assessments \cite{hong2023}.

\subsubsection{High-Sensitivity Long-Term Monitoring with a 3.5-Meter-Class Space Telescope}

To date, many studies of stellar flares and magnetic activity have focused on individual events or relatively limited samples of stars. However, a comprehensive assessment of exoplanet habitability requires long-term observations and statistical characterization of stellar magnetic activity across a wide range of stellar types and ages. Within the 0.2--1.5~$\mu$m baseline, 3.5mST can measure optical/near-ultraviolet flare rates, flare energies, activity indicators, and long-term variability. X-ray/EUV irradiation and stellar-wind properties, which lie outside the direct baseline capability of 3.5mST, would be incorporated through coordinated measurements or constraints from complementary X-ray, ultraviolet, and radio facilities. Together, these data are essential for constructing realistic atmospheric-response models and evaluating the long-term evolution of planetary environments.

For example, nearby potentially habitable planets around M dwarfs span a wide
range of stellar environments. Proxima Centauri b orbits a highly active flare
star \cite{davenport2016}, whereas Ross 128 b orbits a comparatively quiet
M dwarf \cite{bonfils2018}. Comparing such contrasting systems can help
determine how stellar magnetic activity influences atmospheric retention
and long-term planetary habitability \cite{chen2021}. In addition, the young solar-type star EK Draconis serves as an important analogue for studying the early magnetic activity of the Sun. By combining long-term monitoring with atmospheric spectroscopic observations of such systems, a 3.5-meter-class space telescope could help establish the causal relationships between stellar magnetic activity and planetary environmental evolution.

A \textbf{3.5-meter-class space telescope} would provide powerful capabilities for such investigations by enabling precise monitoring of stellar magnetic activity across large and diverse stellar samples. Its high sensitivity and long-duration observing capability would allow systematic measurements of flare occurrence rates, flare energies, magnetic-activity cycles, and other indicators of stellar activity. In particular, by comparing the activity levels of stars hosting planets within their habitable zones with the observed or inferred atmospheric properties of those planets, a 3.5-meter-class space telescope could place robust quantitative constraints on the physical conditions required for planetary habitability. Such observations would contribute significantly to developing a unified understanding of how stellar magnetic environments influence the long-term evolution and potential habitability of exoplanets.

\section{Near-Earth Objects Threatening Earth}\label{near-earth-objects-threatening-earth}

\subsection{Near-Earth Object Surveillance with a 3.5-Meter Space Telescope}\label{near-earth-object-surveillance-with-a-3.5-meter-space-telescope}

The surveillance of \textbf{Near-Earth Objects (NEOs)} is a critical area of space science aimed at ensuring the long-term safety of Earth. NEOs consist primarily of \textbf{asteroids} and \textbf{comets}, some of which have orbits that approach or intersect Earth\textquotesingle s orbit, creating a potential impact hazard. The identification, tracking, and physical characterization of potentially hazardous NEOs are essential for impact-risk assessment and the development of planetary defense strategies. The importance of this research is underscored by evidence of past impact events, such as the Chicxulub impact, which had profound consequences for life on Earth.

NEO surveillance using space-based telescopes offers significant advantages over ground-based observations. Ground observatories are constrained by atmospheric scattering, the day--night cycle, weather conditions, and seasonal visibility limitations, all of which directly affect observational efficiency. In contrast, space-based platforms operate free from these restrictions and can provide more continuous and comprehensive sky coverage. 
For the baseline 3.5-meter space telescope, NEO observations are focused on rapid recovery, precision astrometry, time-resolved photometry, and visible/near-infrared characterization. Thermal-emission measurements required for more direct constraints on NEO sizes and albedos are provided through coordinated observations with ground-based mid-infrared facilities.

\begin{figure}[htbp]
  \centering
  \includegraphics[width=0.90\linewidth,height=0.45\textheight,keepaspectratio]{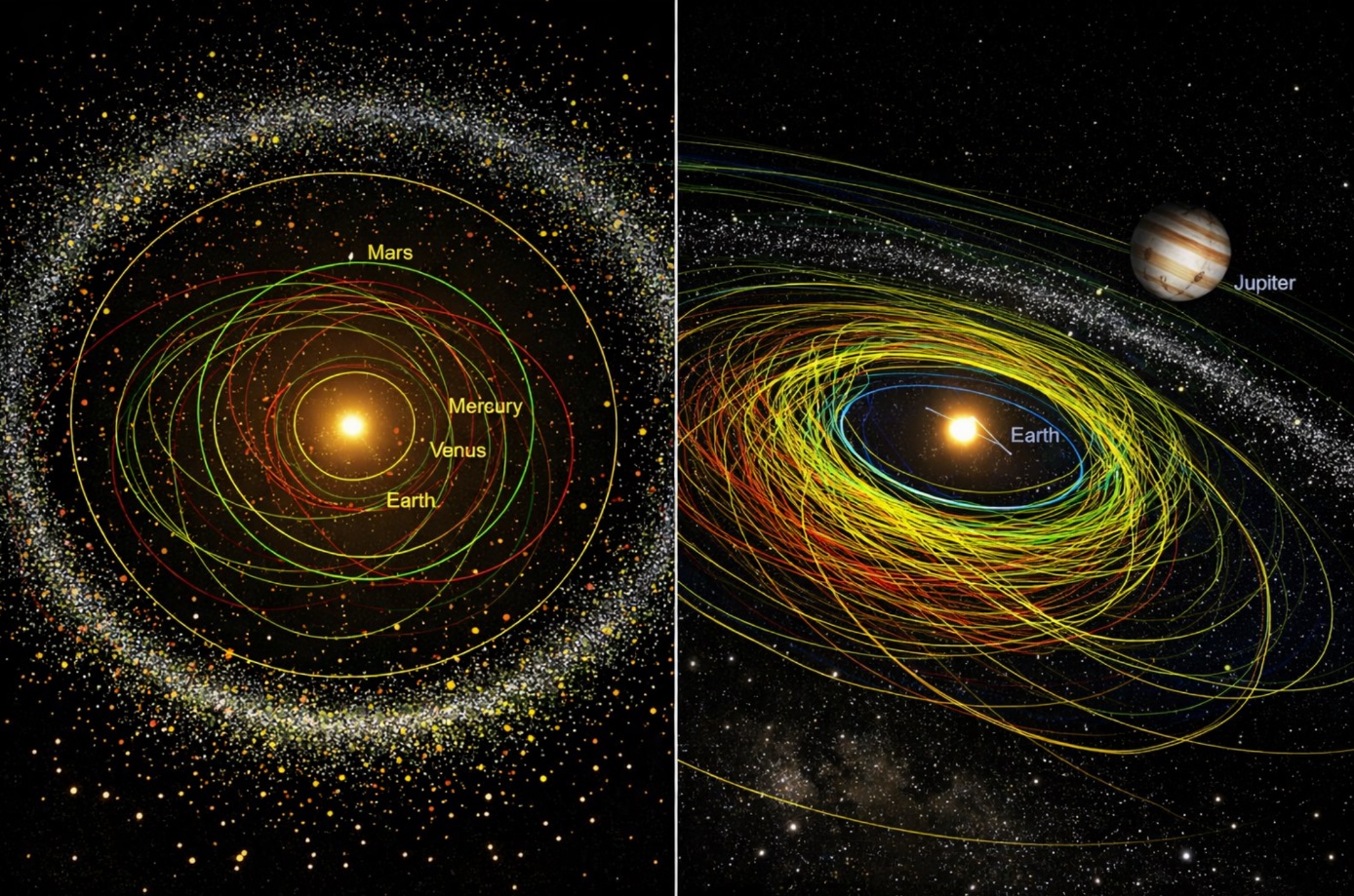}
  \caption{AI-generated conceptual visualization of near-Earth-object orbits; shown for illustration rather than as a quantitative orbital map.}
  \label{fig:neo-orbit-concept}
\end{figure}

\begin{figure}[htbp]
  \centering
  \includegraphics[width=0.90\linewidth,height=0.45\textheight,keepaspectratio]{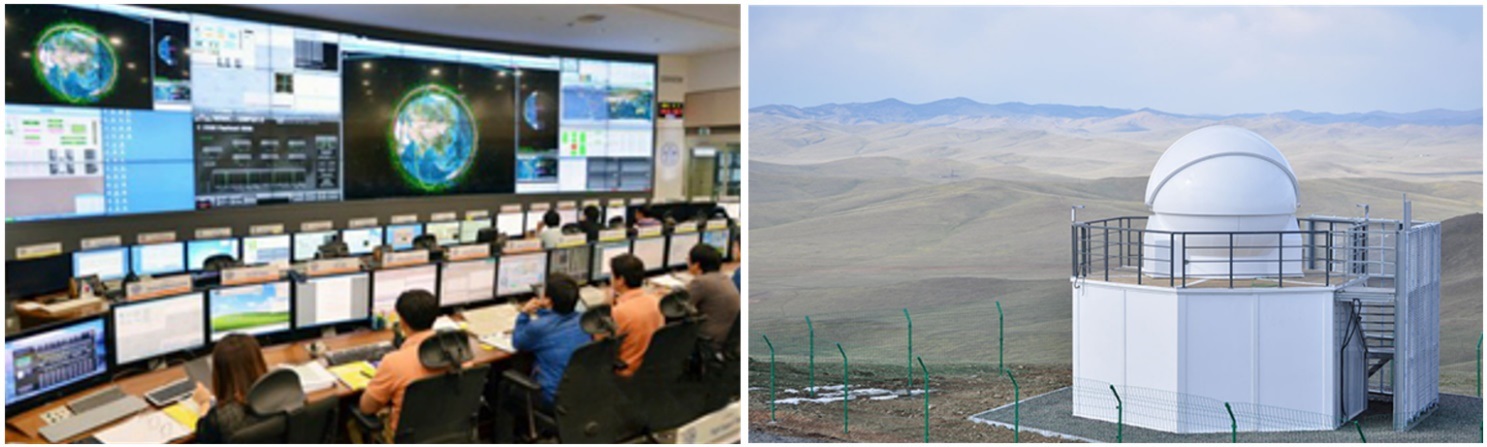}
  \caption{KASI's Optical Wide-field Patrol (OWL) system for space-object surveillance (left) and the OWL observatory in Mongolia (right).}
  \label{fig:owl-surveillance}
\end{figure}

\subsubsection{Scientific Motivation for Near-Earth Object Surveillance}

Accurately determining the orbital and physical properties of potentially hazardous \textbf{Near-Earth Objects (NEOs)}---including their trajectories, sizes, albedos, rotation states, and surface characteristics---is the fundamental first step in assessing impact probabilities and developing effective planetary defense strategies. For NEOs with diameters ranging from several tens to hundreds of meters, the determination of orbital elements alone can often provide an initial estimate of impact risk. However, even much smaller objects, typically in the \textbf{10--30 m size range}, can produce significant regional damage upon atmospheric entry or impact. Consequently, the detection and monitoring of small NEOs remain an important scientific and societal challenge.

The kilometer-scale NEO population is now largely cataloged, whereas the population near the \textbf{140 m} planetary-defense threshold remains substantially incomplete \cite{nasa2023}. The incompleteness becomes still greater at smaller sizes. A notable example is the Chelyabinsk meteor, a roughly 20-meter object that entered Earth's atmosphere in 2013 without prior detection and produced significant regional damage. Such events highlight the continuing need for coordinated NEO discovery, rapid recovery, and characterization programs.

The 3.5-meter space telescope can provide stable visible/near-infrared
photometry, colors, spectra, and precise astrometry of NEOs. These
measurements constrain rotation, taxonomy, and reflected-light properties
but do not by themselves remove the degeneracy between size and albedo.
Combining the reflected-light measurements obtained with 3.5mST with
complementary thermal-infrared observations from ground-based facilities
can provide more reliable constraints on NEO sizes, albedos, and physical
properties.

\subsubsection{Role and Capabilities of a 3.5-Meter Space Telescope}

A \textbf{3.5-meter-class space telescope} can provide next-generation capabilities for space-based Near-Earth Object (NEO) surveillance. Its major contributions include the following:

\subsubsection{Rapid Recovery and Follow-up of Small and Faint NEOs}\label{rapid-recovery-and-follow-up-of-small-neos}

The large collecting area and high sensitivity of the 3.5-meter space
telescope would enable rapid recovery and follow-up of smaller and fainter
NEOs identified by external survey programs and would facilitate early
refinement of their orbital solutions.
Rather than serving as a competitive blind-survey facility, the baseline mission is optimized for rapid recovery and characterization of targets identified by external survey networks. Prompt follow-up observations can extend observational arcs, reduce orbital uncertainties, and support subsequent physical characterization.

\subsubsection{Improved Orbital Accuracy}\label{improved-orbital-accuracy}

Space-based observations are free from atmospheric distortion and seeing effects, resulting in substantially improved astrometric precision. More accurate positional measurements allow for more precise determination of orbital elements and enhance the reliability of long-term orbital predictions. Repeated observations over extended periods further reduce uncertainties in trajectory calculations, thereby improving impact-risk assessments.

\subsubsection{Physical Characterization of NEOs}\label{physical-characterization-of-neos}

Visible and near-infrared photometry and spectroscopy from the 3.5-meter space telescope can constrain NEO colors, taxonomy, rotation, and surface properties. For selected targets, coordinated thermal-emission measurements from ground-based mid-infrared facilities can be combined with these reflected-light observations to improve estimates of physical size and albedo. These physical characteristics are essential for estimating the consequences of potential impacts, understanding surface geology and material properties, and interpreting rotational and dynamical behavior. Such information also supports the design of future mitigation or deflection strategies should a hazardous object be identified.

\subsubsection{Long-Term Monitoring and Expanded Sky Coverage}\label{long-term-monitoring-and-expanded-sky-coverage}

Space-based observations can provide long and repeated observing sequences
without interruptions from weather or the terrestrial day--night cycle,
although the actual observing window remains subject to spacecraft
field-of-regard and scheduling constraints. Such continuity is particularly
useful for rapid target recovery, repeated astrometric measurements, and
time-resolved monitoring of NEOs identified by external survey facilities.

These capabilities would enable a more systematic and quantitative approach to \textbf{Near-Earth Object hazard assessment}. For example, accurate measurements of an NEO\textquotesingle s size, shape, and albedo can significantly improve estimates of impact energy, atmospheric entry behavior, and the geographical extent of potential damage. Consequently, a 3.5-meter-class space telescope would represent a powerful tool for advancing planetary defense and improving humanity's preparedness for future NEO impact threats.

\subsubsection{Technical Requirements for Near-Earth Object Surveillance}

An effective NEO surveillance system requires several key technical capabilities:

\begin{itemize}
\item
  \textbf{Low-stray-light imaging and accurate non-sidereal tracking} --- the ability to measure faint, moving targets against structured backgrounds while maintaining astrometric accuracy and stable image quality.
\item
  \textbf{Rapid target acquisition and flexible scheduling} --- the capability to respond promptly to external NEO alerts and obtain repeated astrometric observations before positional uncertainties grow substantially.
\item
  \textbf{Precise timing and rapid cadence observations} --- highly accurate time measurements and frequent repeated observations are essential for determining and refining the orbits of fast-moving objects.
\item
  \textbf{Visible/NIR spectrophotometric capability} --- sufficient wavelength coverage and calibration stability to constrain NEO colors, taxonomy, rotation, and reflected-light properties. 
\end{itemize}

Complementary thermal-infrared observations play a crucial role in accurately characterizing NEOs. Observations based solely on reflected sunlight suffer from a degeneracy between an object\textquotesingle s size and albedo: two objects with the same apparent brightness may differ significantly in their actual sizes and reflectivities. In contrast, infrared thermal-emission measurements depend directly on an object\textquotesingle s surface temperature and physical size. By combining reflected-light and thermal-emission observations, astronomers can disentangle these effects and derive more reliable estimates of size, albedo, and thermophysical surface properties. Bulk density requires an independent mass constraint in addition to size information.

\subsubsection{Complementarity Between International Surveillance Networks and Space Telescopes}

Current NEO surveillance efforts rely on close cooperation between ground-based survey networks and space-based observatories. A notable example of a completed space-based infrared NEO survey is NEOWISE, while major ground-based programs such as Catalina Sky Survey, Pan-STARRS, and ATLAS continue to discover and track NEOs.

Infrared observations from space provide sensitivity to objects that are difficult to detect from the ground, particularly small and low-albedo bodies. These data complement ground-based observations and serve as valuable calibration and validation references for orbital determination and physical characterization \cite{mainzer2011}.

Ground-based telescopes have the advantage of surveying large regions of the sky rapidly and at relatively low operational cost. However, their performance is inevitably affected by atmospheric conditions, weather, seeing variations, and the day--night cycle. Space-based observations are free from these limitations and benefit from darker backgrounds, longer continuous observing sequences, subject to spacecraft field-of-regard constraints, and stable observing conditions. The combination of ground- and space-based observations therefore represents a highly effective strategy for improving the \textbf{completeness} of NEO surveys and enabling the construction of more comprehensive catalogs that include both large and small potentially hazardous objects.

\subsubsection{NEO Data Processing and Impact Risk Assessment}

Following NEO surveillance observations, the collected data must be incorporated into orbital-dynamics models and impact-probability calculations. Prominent examples include the NASA Jet Propulsion Laboratory \textbf{Sentry} impact-monitoring system and the European Space Agency \textbf{Near-Earth Object Coordination Centre (NEOCC)}, both of which calculate impact probabilities using the orbital elements of detected objects. Integrating NEO observations obtained from space-based telescopes with these monitoring systems can significantly improve the precision and reliability of impact-risk assessments. This demonstrates that NEO surveillance serves not only as a detection activity but also as a critical source of quantitative data for risk analysis, impact forecasting, and the development of mitigation strategies.

\subsubsection{Summary and Future Prospects}

A \textbf{3.5-meter-class space telescope} can make substantial contributions to NEO surveillance through its high sensitivity, broad wavelength coverage, long integration times, and space-based observing environment. These capabilities complement existing ground-based survey networks and are particularly valuable for the detection and characterization of small NEOs that are difficult to identify using current facilities. Beyond providing essential information for planetary defense and impact-risk assessment, such observations also contribute to our understanding of the origin, evolution, and dynamical behavior of small-body populations throughout the Solar System.

\subsection{Space Hazard Analysis of Near-Earth Objects}

\textbf{Near-Earth Objects (NEOs)} are asteroids and comets whose orbits approach or intersect that of Earth. Some of these objects pose potentially significant threats to human society and the terrestrial environment because of the possibility of impact. Consequently, research aimed at protecting Earth from such hazards has evolved beyond simple detection and tracking into a distinct and systematic field known as \textbf{space hazard analysis}.

Space hazard analysis extends beyond identifying the existence and orbital characteristics of NEOs. Its primary objective is to quantitatively evaluate and predict risk by combining the \textbf{probability} of an impact event with its potential \textbf{consequences}. Achieving this goal requires high-precision determination of an object\textquotesingle s physical properties and orbital parameters, modeling of its long-term orbital evolution, rigorous assessment of uncertainties, and the integration of these factors into comprehensive \textbf{risk metrics}. Through this approach, researchers can better characterize potential hazards and support informed decision-making for planetary defense planning and impact-mitigation strategies.

\subsubsection{Space Hazard Analysis: Integrating Probability, Consequences, and Systematic Risk Assessment}

Historically, NEO research has focused primarily on the detection and orbital determination of potentially hazardous objects. However, orbital information alone is insufficient for assessing actual impact risk. Risk cannot be represented by a single value; rather, it must be treated as a function of a probability distribution that combines uncertainties in orbital predictions with outcome-related variables such as an object\textquotesingle s size, density, composition, and impact velocity.

Modern risk-analysis frameworks have developed a variety of tools to quantify the relationship between impact probability and potential consequences. Two of the most widely used approaches are the \textbf{Palermo Technical Impact Hazard Scale} and the \textbf{Torino Scale}. The Palermo Scale is a logarithmic metric designed for specialists, comparing the probability and expected impact energy of a potential collision against the background impact risk. In contrast, the Torino Scale provides a simplified rating system ranging from 0 to 10, enabling the general public and policymakers to more easily understand the significance of a potential impact threat. Through the use of such metrics, the question of \emph{``How dangerous is this object in practice?''} can be addressed in a systematic and quantitative manner rather than relying solely on the fact that an object has been discovered.

Another fundamental aspect of hazard analysis is \textbf{uncertainty propagation}. Initial observations often cover only a short observational arc, resulting in large uncertainties in the estimated orbital elements and their associated covariance matrices. As additional observations are acquired over time, these uncertainties generally decrease. However, when a close approach to Earth is anticipated, nonlinear dynamical effects become increasingly important. Consequently, simple linear covariance propagation is often insufficient. Instead, modern impact-monitoring systems employ methods that sample ensembles of possible orbital solutions throughout orbital-parameter space and search for \textbf{virtual impactors}---orbital solutions that lead to a future collision with Earth \cite{milani2005}.

The combined evaluation of impact probability, impact consequences, and uncertainty propagation forms the foundation of quantitative space hazard analysis. Nevertheless, two major challenges remain. First, the distributions of impact consequences must be refined through improved knowledge of an object\textquotesingle s physical properties, including its size, density, shape, and albedo. Second, comprehensive dynamical models must incorporate \textbf{non-gravitational perturbations}, such as thermal forces and the \textbf{Yarkovsky effect}, which can gradually alter an object\textquotesingle s orbit over long timescales and significantly affect long-term impact predictions. Addressing these challenges is essential for improving the accuracy and reliability of future NEO hazard assessments.

\subsubsection{Physical Properties of NEOs and Impact Consequences: The Importance of Separating Albedo and Size}

Traditionally, the size of a \textbf{Near-Earth Object (NEO)} is estimated from its \textbf{absolute magnitude (H)} and \textbf{albedo}. However, optical observations alone cannot uniquely determine these two quantities because the observed brightness depends on both the object\textquotesingle s size and its reflectivity. As a result, uncertainties in albedo translate directly into uncertainties in size estimates. Since the size of an NEO is a primary parameter in determining its impact energy (approximately proportional to \textbf{$\frac{1}{2}mv^2$}), accurate size determination is essential for reliable assessments of potential impact consequences.

To overcome this limitation, observations of an NEO's \textbf{thermal emission} are required. Infrared measurements provide direct information about an object\textquotesingle s thermal radiation, allowing estimates of its surface temperature and physical size. When combined with reflected-light observations, thermal-emission data help break the degeneracy between size and albedo, enabling more accurate determination of both parameters \cite{mainzer2011}. 
In the baseline mission concept, these thermal-emission measurements are not obtained by the 3.5-meter space telescope itself but through coordinated observations with ground-based mid-infrared facilities.
This combined optical--infrared approach allows the estimation of absolute sizes and physical classifications that cannot be achieved through optical observations alone, thereby reducing uncertainties in impact-consequence models.

Furthermore, spectroscopic observations of reflected light can provide information on an object\textquotesingle s composition---such as whether it is dominated by silicates, carbonaceous materials, or metallic constituents---as well as its surface physical properties, including scattering behavior and surface roughness. These characteristics influence fragmentation processes, atmospheric entry dynamics, and energy-transfer mechanisms during an impact event. Consequently, they are critical inputs for \textbf{sensitivity analyses} within quantitative hazard-assessment models and for improving predictions of potential damage scenarios.

\subsubsection{Non-Gravitational Perturbations: The Yarkovsky Effect and Long-Term Risk Evolution}

One of the greatest challenges in long-term hazard prediction is accounting for \textbf{non-gravitational perturbations} that alter the orbits of NEOs over extended periods. Among the most important of these effects is the \textbf{Yarkovsky effect}, a mechanism in which absorbed solar radiation is re-emitted as thermal radiation, producing a small but continuous force that gradually modifies an asteroid's orbit.

Although the resulting acceleration is extremely weak, its cumulative influence can become significant over timescales of decades or longer, particularly for small NEOs. In some cases, the Yarkovsky effect can alter orbital trajectories sufficiently to change future close-approach geometries and significantly modify predicted impact probabilities.

Numerical studies have demonstrated that neglecting the Yarkovsky effect may lead to substantial inaccuracies in long-term impact-risk assessments \cite{vokrouhlicky2015}. Consequently, modern hazard-analysis frameworks increasingly incorporate Yarkovsky-related parameters or plausible ranges of thermal-force models into \textbf{Monte Carlo simulations} and long-term orbital-evolution calculations. By doing so, researchers can explore a broader range of possible future trajectories and obtain a more realistic assessment of impact probabilities than would be possible using orbital covariance propagation alone. Such approaches provide a deeper understanding of long-term NEO risk evolution and improve the reliability of future planetary-defense planning.

\subsubsection{The Role of a 3.5-Meter Space Telescope: From Surveillance to Hazard Analysis and Mitigation Planning}

A \textbf{3.5-meter-class space telescope} has the potential to significantly strengthen multiple stages of space hazard analysis. Its contributions extend well beyond simple object detection and can be summarized as follows:

\subsubsection{1) Improving the Completeness of Early Detection}\label{improving-the-completeness-of-early-detection}

The 3.5-meter space telescope can contribute to the recovery and
characterization of small NEOs identified by external survey programs
through sensitive visible and near-infrared imaging and precise astrometry.
Its large collecting area and stable observing environment would be
particularly valuable for faint and rapidly moving objects that require
prompt follow-up before their positional uncertainties increase
substantially. Thermal-emission measurements needed to determine physical
size and albedo would be obtained through coordinated observations with
ground-based mid-infrared facilities rather than by the 3.5-meter telescope
itself.

\subsubsection{2) Reducing Orbital Uncertainties}\label{reducing-orbital-uncertainties}

Orbital tracking relies on repeated \textbf{astrometric measurements} to refine an object\textquotesingle s trajectory and reduce uncertainties. The stable observing environment and long-term monitoring capability of a space telescope can significantly decrease the covariance associated with orbital solutions. As uncertainties shrink, families of possible orbital trajectories converge toward a smaller set of solutions, allowing potential impact scenarios to be confirmed or eliminated at much earlier stages. This improvement directly enhances the reliability of long-term impact predictions.

\subsubsection{3) Constraining Physical Properties}\label{constraining-physical-properties}

Visible and near-infrared photometry and spectroscopy with the 3.5-meter
space telescope can constrain NEO colors, spectral type, rotation, and
surface composition. These measurements can be combined with thermal-emission
data obtained from coordinated ground-based mid-infrared observations to
improve estimates of physical size and albedo. The resulting joint dataset
provides more reliable physical parameters for impact-consequence
calculations and hazard-assessment models, while reflected-light
spectroscopy from the space telescope supplies additional information on
surface composition and material properties.

\subsubsection{4) Constraining Non-Gravitational Perturbations}\label{constraining-non-gravitational-perturbations}

Non-gravitational effects, such as the \textbf{Yarkovsky effect}, depend strongly on an object\textquotesingle s rotational state and thermal properties. 
Through repeated photometric monitoring, the 3.5-meter space telescope
could constrain rotation periods, spin states, and light-curve properties.
When combined with complementary thermal-infrared measurements, these
observations would provide important inputs for thermophysical models and
for simulations of Yarkovsky-driven orbital evolution.
As a result, future trajectory predictions and impact-risk assessments become significantly more reliable.

In this way, a 3.5-meter-class space telescope would transform NEO research from a primarily detection-oriented effort into a comprehensive framework encompassing discovery, characterization, hazard analysis, and ultimately the development of informed planetary-defense and mitigation strategies.

\subsubsection{5) Providing Information for Policy and Mitigation Strategies}\label{providing-information-for-policy-and-mitigation-strategies}

The results of high-precision hazard analyses extend beyond the realm of scientific research and serve as essential inputs for policymakers and emergency-planning authorities. Accurate risk assessments enable decision-makers to establish priorities for impact mitigation, allocate resources efficiently, and develop appropriate response strategies when necessary. For example, objects assigned elevated ratings on the \textbf{Palermo Technical Impact Hazard Scale} or the \textbf{Torino Scale} may warrant immediate high-precision follow-up observations and the initiation of detailed studies on potential mitigation measures.

Taken together, these capabilities demonstrate that a \textbf{3.5-meter-class space telescope} would contribute not only to the recovery and characterization of a larger number of NEOs but also to the strengthening of a comprehensive hazard-assessment framework encompassing \textbf{detection, characterization, risk evaluation, and mitigation planning}.

\subsection{Predictive and Early Warning Analysis for Space Hazards}

Research on the detection and tracking of Near-Earth Objects (NEOs) and Potentially Hazardous Objects (PHOs) is closely connected to the broader scientific and societal objective of ensuring the long-term safety and sustainability of human civilization and the terrestrial environment. While traditional NEO studies have focused primarily on the orbital dynamics of individual objects, predictive and early warning analysis for space hazards emphasizes the early identification of potential future threats, the prediction and evaluation of risk scenarios, and the establishment of effective warning systems.

This field integrates not only information about individually detected objects but also the statistical properties of the overall NEO population, the temporal evolution of hazard events, and comprehensive modeling that combines orbital and physical characteristics. By incorporating these diverse sources of information, predictive and early warning systems aim to provide timely assessments of potential impact risks and support informed decision-making for planetary defense.

The following sections discuss the fundamental concepts of space-hazard prediction and early warning, the principal analytical and computational techniques employed in this field, the role and expected performance improvements enabled by a 3.5-meter-class space telescope, and the quantitative characteristics of such research as demonstrated through relevant studies and references.

\subsubsection{The Need for Space Hazard Prediction and Early Warning Systems}

Potentially hazardous objects often possess significant orbital uncertainties at the time of their discovery. Initial observations typically cover only a \textbf{short observational arc}, making long-term orbital predictions subject to substantial uncertainty. Consequently, there is a critical need for \textbf{predictive and early warning systems} capable of identifying whether a Near-Earth Object (NEO) may pose an impact threat at some future time and of recognizing potentially hazardous scenarios as early as possible.

An effective warning system must provide a comprehensive response framework for NEO candidates whose estimated risk exceeds a specified threshold. Such a framework includes immediate and continuous follow-up observations, high-precision orbital calculations, consideration of non-gravitational perturbations, and ongoing evaluation of how impact probabilities evolve over time. Widely used hazard indicators, such as the \textbf{Palermo Technical Impact Hazard Scale} and the \textbf{Torino Scale}, are employed to quantify the severity of potential impact threats. Objects assigned Palermo Scale values above zero generally warrant additional expert review and long-term monitoring.

Importantly, an early warning system should do more than simply assign a risk rating. It must be designed to predict how impact risk may evolve in the future and to support the optimization of mitigation and response strategies as new information becomes available. A central objective is therefore the time-dependent modeling of how uncertainties in orbital elements and physical properties influence hazard metrics and risk assessments.

\subsubsection{Predictive and Early Warning Methodologies}

\subsubsection{1) Long-Term Orbital Prediction Modeling}\label{long-term-orbital-prediction-modeling}

Reliable long-term prediction of NEO trajectories requires consideration not only of gravitational forces but also of \textbf{non-gravitational perturbations}, particularly the \textbf{Yarkovsky effect}. The Yarkovsky effect arises when solar radiation heats an object\textquotesingle s surface and the absorbed energy is subsequently re-emitted as thermal radiation, generating a small but persistent force. Although subtle, this force can accumulate over long periods and significantly alter the trajectories of small NEOs. As a result, it represents one of the principal sources of uncertainty in long-term impact predictions and must be incorporated into modern hazard-assessment models.

Numerous studies have demonstrated that multivariate \textbf{Monte Carlo simulations} incorporating the Yarkovsky effect can substantially improve the accuracy of long-term impact-risk forecasts \cite{vokrouhlicky2015}. Such simulations require a nonlinear probabilistic framework that simultaneously accounts for uncertainties in the orbital \textbf{state vector}, non-gravitational perturbation parameters, observational errors, and the temporal spacing of observations.

A common approach involves generating large ensembles of \textbf{virtual clones} representing plausible orbital solutions within the uncertainty region of the observed orbit. Each clone is then propagated forward in time to determine its future trajectory and minimum Earth-approach distance. By analyzing the statistical behavior of these orbital ensembles, researchers can estimate the time evolution of impact probabilities and identify periods of elevated risk. This methodology typically follows the sequence:

\textbf{Input Uncertainties $\rightarrow$ Impact Probability Distribution $\rightarrow$ Prediction of Warning and Risk-Escalation Timelines}

Such an approach provides a quantitative foundation for future early warning systems and enables more reliable long-term forecasting of potential NEO impact hazards.

\subsubsection{2) Integrated Risk Scales and Warning Criteria}\label{integrated-risk-scales-and-warning-criteria}

The \textbf{Palermo Technical Impact Hazard Scale} and the \textbf{Torino Scale} are the most widely used tools for impact-risk assessment. The Palermo Scale expresses risk as a logarithmic measure that combines impact probability with expected impact energy, facilitating quantitative comparisons among potential threats by specialists. The Torino Scale, in contrast, is designed for public communication and provides a simplified classification system that combines impact probability and potential consequences into a single risk level.

For example, objects assigned a \textbf{Torino Scale value of 2 or higher} generally warrant additional observations and are often interpreted as representing a temporary warning level. Objects reaching \textbf{Torino Scale values of 5 or higher} are typically considered significant enough to justify substantial scientific attention and preparedness planning. Together, these scales provide a common framework for translating complex impact-risk assessments into standardized warning indicators.

\subsubsection{Role of a 3.5-Meter Space Telescope and Expected Performance Improvements}

A \textbf{3.5-meter-class space telescope} has the potential to substantially enhance multiple stages of predictive and early-warning research for space hazards. Its major contributions can be summarized as follows:

\subsubsection{(A) Improved Detection and Reduction of Initial Uncertainties}\label{a-improved-detection-and-reduction-of-initial-uncertainties}

The large collecting area and high sensitivity of a space telescope can support rapid recovery and characterization of smaller and fainter NEOs discovered by external survey programs, facilitating early refinement of their orbital solutions. The positional uncertainties associated with short observational arcs immediately after discovery can be significantly reduced through additional high-precision astrometric observations. This improvement directly reduces the initial uncertainty space used in long-term probability simulations and impact-risk assessments \cite{milani2005}.

\subsubsection{(B) Expanded Physical Characterization}\label{b-expanded-physical-characterization}

By combining visible/near-infrared observations from 3.5mST with thermal-infrared measurements from coordinated ground-based facilities, the size, albedo, and spectral properties of an NEO can be constrained more reliably. These parameters constitute critical inputs for impact-consequence models. In particular, the combined analysis of reflected and thermal radiation breaks the degeneracy between size and albedo, significantly reducing uncertainties in estimates of impact energy and potential damage \cite{mainzer2011}.

\subsubsection{(C) Improved Long-Term Orbital Predictions}\label{c-improved-long-term-orbital-predictions}

Long-term monitoring over periods of several years can substantially improve
orbit determination. When combined with archival and external observations,
the effective observational baseline can be extended to substantially longer
timescales.

\subsubsection{(D) Identification of Critical Warning Thresholds}\label{d-identification-of-critical-warning-thresholds}

One of the most important objectives of an early-warning system is determining when a risk metric exceeds a predefined threshold. By providing highly accurate observations that constrain the initial conditions of long-term orbital simulations, a 3.5-meter-class space telescope can identify such threshold crossings at earlier stages. Earlier detection of elevated risk levels increases the time available for scientific assessment, policy development, and potential mitigation planning.

\subsubsection{(E) Multi-Signal Observations and Hazard Information Integration}\label{e-multi-signal-observations-and-hazard-information-integration}

Observations obtained by a 3.5-meter-class space telescope can be integrated with ground-based optical and radar measurements as well as data from other space missions, such as the James Webb Space Telescope. Such multi-platform data fusion can substantially improve the accuracy of impact-risk assessments. For example, combining radar-derived distance and velocity measurements with space-based physical characterization data yields more precise orbital solutions and more reliable long-term predictions.

\subsubsection{Expected Outcomes and Applications}

Predictive and early-warning research for space hazards provides benefits that extend far beyond the issuance of impact alerts. Comprehensive time-dependent risk models can forecast the evolution of impact probabilities over periods spanning decades and identify when and where critical risk thresholds may emerge. These predictive risk maps can serve as valuable decision-support tools for public policy, emergency management, and disaster-preparedness planning.

In addition, long-term observational datasets contribute to studies of \textbf{NEO population statistics} and the overall structure of small-body populations in the Solar System. By revealing the spatial and temporal distributions of cumulative risk, these datasets improve our understanding of how hazardous-object populations evolve and help refine the methodologies used to estimate impact risk. Ultimately, such research contributes not only to planetary defense but also to the broader scientific understanding of the dynamical evolution of near-Earth object populations.

\chapter{Conclusion}

The 3.5-meter Segmented-Mirror Robotic Space Telescope is being studied as
a large-aperture space telescope capable of supporting a broad range of
astrophysical investigations from the optical to the near-infrared. Its
baseline concept combines a 3.5-meter segmented primary mirror with
wide-field imaging, spectroscopy, and a coronagraphic capability, providing
a common observational platform for both survey-oriented and targeted
science programs.

The scientific program spans several complementary areas, including
cosmology and galaxy evolution, compact-object astrophysics, time-domain
and multi-messenger astronomy, direct imaging and characterization of
nearby planetary systems, and observations of Solar-System small bodies.
These programs place different demands on sensitivity, angular resolution,
spectral resolution, photometric stability, observing cadence, and
high-contrast performance. The mission architecture is therefore being
developed by considering these science requirements together rather than
optimizing the telescope for a single experiment.

The current telescope concept adopts a segmented aperture to obtain the
collecting area and angular resolution of a 3.5-meter-class telescope
within realistic launch constraints. Wide-field imaging is intended to
support deep surveys, precision photometry, and repeated monitoring, while
the spectroscopic system is being studied with a baseline resolving power
of approximately $R \sim 1000$ and higher-resolution options approaching
$R \sim 5000$ for selected science applications. A dedicated coronagraphic
capability is also included in the science concept to enable investigations
of nearby exoplanetary systems and to establish experience in high-contrast
observations with a segmented space telescope.

Several elements of the mission remain the subject of continuing trade
studies. These include the detailed instrument configuration, extension of
the near-infrared wavelength coverage, the final mission orbit, detector
selection, wavefront-control performance, and the balance between survey
and targeted observing programs. Refinement of these elements would require
more detailed science-performance simulations and engineering studies as
the mission concept advances.

Beyond its immediate scientific objectives, the proposed telescope would
provide an opportunity to contribute to building national capability in the
integrated development and operation of large-aperture space telescopes.
Experience with segmented-mirror optics, precision wavefront control, space-based
scientific instrumentation, and long-term astronomical observations would
provide a foundation for future Korean participation in larger international
space-astronomy missions.

The present white paper defines an initial science and architecture concept
for the 3.5-meter Segmented-Mirror Robotic Space Telescope. Continued
development of the concept would focus on translating the scientific
objectives presented here into quantitative instrument and system
requirements and on evaluating their feasibility through detailed mission
and technology studies.

\chapter*{References}\addcontentsline{toc}{chapter}{References}

% The document class used by the white-paper template generates an additional
% section/chapter heading inside thebibliography. Suppress that local heading
% so that only the chapter-level ``References'' heading above is printed.
\begingroup
\makeatletter
\renewcommand{\section}{\@ifstar\@gobble\@gobble}
\renewcommand{\chapter}{\@ifstar\@gobble\@gobble}
\makeatother

\endgroup

%\appendix
%\renewcommand{\chaptername}{Key Science}   % chapter head reads "Key Sceince A/B"
%\renewcommand{\appendixname}{Key Science}

% per-appendix banner/section colour theme (main body stays blue), muted tones.
% \clearpage before each switch ships the previous page with its own colour
% (otherwise \include's own \clearpage would recolour the preceding page).

%\clearpage
%\definecolor{Blue1}{HTML}{9488AD}\definecolor{Blue2}{HTML}{6F6295}\definecolor{Blue3}{HTML}{514670}% Appendix A: muted violet
%\include{appendix_A}
%\clearpage
%\definecolor{Blue1}{HTML}{86A67C}\definecolor{Blue2}{HTML}{5E8158}\definecolor{Blue3}{HTML}{45613F}% Appendix B: muted sage green
%\include{appendix_B}
%\clearpage
%\definecolor{Blue1}{HTML}{C89562}\definecolor{Blue2}{HTML}{A4612E}\definecolor{Blue3}{HTML}{713F24}% Appendix C: muted amber and burnt orange
%\include{appendix_C_book}
%\clearpage
%\definecolor{Blue1}{HTML}{D98A9A}\definecolor{Blue2}{HTML}{B34F68}\definecolor{Blue3}{HTML}{7A2944}% Appendix D: muted rose and crimson
%\include{appendix_X_book}
\clearpage
\definecolor{Blue1}{HTML}{1FABD5}\definecolor{Blue2}{HTML}{1D8DB0}\definecolor{Blue3}{HTML}{116E8A}% reset to blue for the back cover

\backmatter
\MakeBackCover

\end{document}